\documentclass[aps, prd, showpacs, twocolumn, nofootinbib, preprintnumbers, floatfix,letterpaper]{revtex4-1}
\usepackage[utf8]{inputenc}
\usepackage{bm}
\usepackage{latexsym,amssymb,amsmath,amsfonts}
\usepackage{graphicx}
\usepackage{longtable}

\usepackage{bm}
\usepackage{footnotebackref}

\usepackage[usenames]{xcolor}
\usepackage{soul}
\usepackage{makecell}
\usepackage{graphicx}
\usepackage{amssymb}
\usepackage{amsmath}
\usepackage{longtable}
\usepackage{rotating}
\usepackage{color}
\usepackage{fancyhdr}
\usepackage{ifthen}
\usepackage{slashed}
\usepackage{xspace}
\usepackage{tabularx}
\usepackage{float}

\def \tsid {t_\mathrm{s}}

\def \hpx{{\tt HEALPix} }

\begin{document}

\title{Unified mapmaking for an anisotropic stochastic gravitational wave background}
\author{Jishnu Suresh}
\email{jishnu@icrr.u-tokyo.ac.jp}
\affiliation{Institute for Cosmic Ray Research (ICRR), The University of Tokyo,
Kashiwa City, Chiba 277-8582, Japan} 
\author{Anirban Ain}
\email{anirban.ain@pi.infn.it}
\affiliation{Istituto Nazionale di Fisica Nucleare (INFN) sezione Pisa, 56126 Pisa, Italy }
\author{Sanjit Mitra}
\email{sanjit@iucaa.in}
\affiliation{Inter-University Centre for Astronomy and Astrophysics (IUCAA), Pune 411007, India}
\begin{abstract}
A stochastic gravitational wave background (SGWB), created by the superposition of signals from unresolved astrophysical sources, may be detected in the next few years. Several theoretical predictions are being made about the possible nature of anisotropies in the background. Estimating the variation of intensity across the sky can, therefore, play a key role in improving our understanding of astrophysical models. Sky maps have been produced for all the data-taking runs of the advanced ground-based interferometric detectors. While these maps are being produced in pixel and spherical harmonic (SpH) bases, to probe, respectively, localized and diffuse astrophysical and cosmological sources, with algorithms that employ cross-correlation as the common strategy, the underlying algebra and numerical implementation remain different. As a consequence, there was a need for producing sky maps in both bases in those analyses. We show that these manifestly redundant
methods could indeed be unified to a single analysis that can probe very different scales and demonstrate it by applying them on real data. We first develop the algebra to show that the results in two different bases are easily transformable. We then incorporate both the schemes in the now-standard analysis pipeline for anisotropic SGWB, {\tt PyStoch}. This will enable SGWB anisotropy searches in SpH basis also to take full advantage of integrated \hpx tools and makes it computationally feasible to perform the search in every frequency bin. We, however, follow a different approach for direct estimation of the SpH moments. We show that the results obtained from these different methods match very well; the differences are less than $0.1$\% for the SpH moments and less than $0.01$\% for the Fisher information matrices. Thus we conclude that a single sky map will be sufficient to describe the anisotropies in a stochastic background.The multiple capabilities of {\tt PyStoch} will be useful for estimating and constraining various measures that characterize an anisotropic background.
\end{abstract}

\maketitle
\section{Introduction}
A new era of astronomy began with the detection of gravitational waves (GWs)~\cite{GW150914} by the Advanced Laser Interferometric Gravitational-wave Observatory (LIGO), followed by tens of binary mergers~\cite{GWTC-2}. These detections have opened up new avenues for exploring the late and early stages of the Universe. A vigorous global effort is underway to observe GW signals in widely separated frequency bands~\cite{ET,CE,LISA}.
Among these sources, the stochastic gravitational wave background (SGWB)~\cite{1997rggr.conf..373A,AllenOttewill} is one of the most interesting ones. A large number of unresolved distant compact binary coalescences~\cite{2018PhRvL.120i1101A,2011PhRvD..84h4004R, 2011PhRvD..84l4037M, 2011ApJ73986Z, 2012PhRvD..85j4024W, 2013MNRAS.431..882Z,2016PhRvD..94j3011D, 2021PhRvD.103d3002P} and millisecond pulsars in galaxy clusters~\cite{hotspot,HUGHES201486,2018arXiv180710620C,PhysRevD.88.062005} can produce an astrophysical background, which are close to the sensitivity levels of the current or upcoming ground-based interferometers. Astrophysical SGWB is one of the primary targets for the network of current-generation ground-based detectors, consisting of LIGO, Virgo~\cite{virgo}, KAGRA~\cite{Kagra} and upcoming LIGO-India~\cite{ligo_india}. The background can be significantly anisotropic due to the nonuniform distribution of astrophysical sources in the local Universe~\cite{PhysRevD.98.063509, 2014PhRvD..89h4076M, 2018PhRvD..98f3501J, 2011MNRAS.411.2549M, 2012PhRvD..86j4007R, 2013PhRvD..87d2002W, 2013PhRvD..87f3004L}. Based on the recent rate estimation from the observed GW signals ~\cite{2018PhRvL.120i1101A} and the pulsar timing array data~\cite{Arzoumanian_2018,NANOgrav}, the detection of SGWB seems promising in the near future. 

Several methods~\cite{Romano2017,Michelson87,christ92,flan93,AllenOttewill,allen01,LazzariniRomano,Alonso_Contaldi} have been proposed in the past to estimate the intensity variation of a SGWB across the sky. This procedure of obtaining the spatial distribution of intensity---the sky map---is often referred to as the mapmaking process. The standard algorithms have been thoroughly investigated and implemented in the pixel basis~\cite{ballmer06,Mitra07}, as well as in the spherical harmonic basis~\cite{Thrane09}.
The pixel basis is suited for localized pointlike sources, like the ``hot spot'' produced by a large number of monochromatic sources in a galaxy cluster~\cite{hotspot, fermiLAT}, whereas the spherical harmonic basis is appropriate for constraining astrophysical parameters by putting upper limits on smooth and diffuse backgrounds produced by, for example, anisotropic distribution of binary mergers in the local Universe~\cite{2018PhRvD..98f3501J, PhysRevD.98.063509, Cusin:2017fwz, 2018arXiv180710620C}, or millisecond pulsars in our Galaxy~\cite{Talukder:2010yd}.
Alternative approaches for the spherical harmonic basis have also been proposed by ~\citet{AllenOttewill,Renzini}. The efficiency of these ``radiometer'' algorithms was dramatically improved through the mechanism of data folding~\cite{Ain_Folding}. Recently, we developed a new pipeline called {\tt PyStoch}~\cite{pystoch,O3-BBR}, which can boost the efficiency of the mapmaking process by another factor of few tens. {\tt PyStoch}  is the first implementation of a radiometer search that is fully integrated with the Hierarchical Equal Area isoLatitude Pixelization of a sphere \hpx~\cite{HEALPix} scheme. \hpx offers highly efficient tools for Fourier transforms on the sky, and other methods useful for sky map analysis and manipulation. Other than the speed improvement in the analysis, we designed the pipeline in a way that the intermediate results, the sky maps at each frequency bin (narrow-band maps), combining data from multiple detectors can all be done straightforwardly in an integrated way. Folded data and {\tt PyStoch} were recently used for SGWB anisotropy analysis by the LIGO-Virgo-KAGRA Collaboration~\cite{O3-BBR}.

All the SGWB anisotropy analyses so far that estimated the spherical harmonic (SpH) moments of the sky, also produced a separate sky map in addition to the pixel-based one. This is because the pixel and SpH domain analyses used different mathematical framework. Here we first show algebraically how the SpH coefficients of the sky map could be precisely estimated from the pixel domain radiometer sky map, instead of running a separate radiometer search in the SpH basis. However, we also propose an alternative method to directly obtain the SpH results, reusing some of the quantities already computed for the pixel domain analysis. Either way, the searches get unified in the same analysis framework, making a single map sufficient to describe the anisotropic sky, and both methods are computationally fast.
The latter method makes it easier to evaluate the Fisher information matrix in the SpH domain, which is necessary to estimate the significance of the results and to put upper limits on the anisotropy. We incorporate both the schemes in the {\tt PyStoch} pipeline enhancing its capabilities, thereby making {\tt PyStoch} the one single code to perform all the current SGWB anisotropy analyses in an efficient way integrated with the powerful \hpx tools.

With this unified mapmaking pipeline, a separate analysis to search for SGWB anisotropy in the SpH domain becomes redundant. Moreover, it will now be possible, without demanding an unreasonable amount of computing resources, to perform the SpH search at every frequency bin, similar to the pixel basis, which became feasible with the introduction of {\tt PyStoch}.
This has the added advantage that, in order to perform a broadband search for different spectral indices, running separate analyses is not necessary; the narrow-band SpH coefficients can be combined with proper weights to obtain the intended results.
A broadband search has very little sensitivity to narrow-band sources, as it averages over all the frequencies which may be adding only noise. Therefore, if a diffuse narrow-band stochastic background is present, a narrow-band SpH search will have much more sensitivity to detect it or to constraint it.
Conducting the analysis for different bases, frequency ranges and spectral shapes is important to search for unknown persistent sources, which is one of the primary goals of GW astronomy, and GW radiometer analyses are optimal for this purpose.

This paper is organized as follows: Sec.~\ref{sec:method} provides a detailed description of the analysis in the spherical harmonic basis starting with a brief review of the GW radiometer algebra. In Sec. \ref{sec:implementation}, we discuss the implementation of the new method in the existing {\tt PyStoch} pipeline and show that the transformed results in the pixel and spherical harmonic bases match very well. In Sec. \ref{sec:conclusion}, we discuss the implications of our result and its immediate advantages for SGWB searches. 

%----------------------------------------------------------------------------
\section{Method}
\label{sec:method}
An SGWB is estimated from the cross-power spectral density (CSD) of data from pairs of detectors~\cite{Romano2017}. Assuming the SGWB frequency spectral shape to be $H(f)$ and that it is the same in every direction on the sky, a search for a specific spectral distribution boils down to estimation of the SGWB sky map $\mathcal{P}(\mathbf{\widehat{\Omega}})$. The ``sky map'' $\mathcal{P}(\mathbf{\widehat{\Omega}})$ is proportional to the flux coming from different directions on the sky~\cite{Mitra07}. One can perform the search in any set of bases $e(\mathbf{\widehat{\Omega}})$ on the two-sphere~\cite{Thrane09}, in which the anisotropy map can be expanded as,
\begin{equation}
\label{POmega:general_basis}
\mathcal{P}(\mathbf{\hat{\Omega}}) \ := \ \sum_p \mathcal{P}_p  e_p(\mathbf{\hat{\Omega}})\, .
\end{equation}
The maximum-likelihood (ML) method for mapping this GW intensity has been implemented in two natural bases, the pixel basis~\cite{Mitra07,ballmer06} and the spherical harmonic basis~\cite{Thrane09}. Even though the search should determine the choice of basis, earlier analyses were forced to choose the SpH basis for the most part~\cite{O1directional,O2directional}, due to severe limitations in computation, in both cost and numerical accuracy, especially in dealing with large Fisher matrices in the pixel basis. Recently, the introduction of {\tt PyStoch} with folded data has made the searches very efficient, which allows us to choose the basis freely, eliminating computational limitations. Initially, {\tt PyStoch} was written for the pixel basis. Here, we show how we enhance {\tt PyStoch} by adding the capability to perform an SpH search, unifying the approaches for the two bases.

In the pixel basis, which is appropriate to search for an anisotropic SGWB dominated by a localized pointlike source, one can decompose the angular power distribution from Eq.~(\ref{POmega:general_basis}) as
\begin{equation}
\mathcal{P}(\mathbf{\hat{\Omega}}) \ = \ \mathcal{P}_{\mathbf{\hat{\Omega}'}} \delta(\mathbf{\hat{\Omega}},\mathbf{\hat{\Omega}'})\,.
\label{POmega:pixel_basis}
\end{equation}
In contrast, the SpH basis is suitable to search for a diffuse background which may be dominated by, say, a dipolar or a quadrupolar distribution. We can expand the anisotropy map over the basis functions $Y_{lm}$ as, 
\begin{equation}
\mathcal{P}(\mathbf{\hat{\Omega}}) \ = \ \sum_{lm} \mathcal{P}_{lm} Y_{lm}(\mathbf{\hat{\Omega}}) \, .
\label{POmega:spherical_basis}
\end{equation}
Here, we follow $Y_{lm}(\theta,\phi)$ conventions used by~\citet{jackson}. A standard ML solution for $\mathcal{\hat P}(\mathbf{\widehat{\Omega}})$ in a general basis, which produces the estimates for the SGWB sky maps, can be obtained using the existing methods as~\cite{Mitra07,Thrane09,pystoch},
\begin{equation}
\label{eq_ML_solution}
 \mathcal{\hat{P}}_{p} \ \equiv \ \hat{\bm{\mathcal{P}}} \ = \ \mathbf{\Gamma}^{-1} \cdot \mathbf{X} \, ,
%\hat{\bm{\mathcal{P}}} \ = \ \mathbf{\Gamma}^{-1} \cdot \mathbf{X} \, ,
\end{equation}
where $\mathbf{X} \equiv X_p$, the ``dirty'' map, is given as
\begin{equation}
\label{eq_dirtymap}
X_p =\frac{4}{\tau} \sum_{Ift} \frac{ H(f) \gamma^{I*}_{ft,p}} {P_{\mathcal{I}_1}(t;f) P_{\mathcal{I}_2}(t;f)} \widetilde{s}_{\mathcal{I}_1}^*(t;f) \widetilde{s}_{\mathcal{I}_2}(t;f) \, ,
\end{equation}
and $\mathbf{\Gamma} \equiv \Gamma_{pp'}$, the Fisher information matrix, as
\begin{equation}
\label{eq_fisher}
\Gamma_{pp'} = 4 \sum_{Ift} \frac{H^2(f)}{P_{\mathcal{I}_1}(t;f) \, P_{\mathcal{I}_2}(t;f)} \,\gamma^{I*}_{ft,p} \, \gamma^{I}_{ft,p'} \, .
\end{equation}
$P_{\mathcal{I}_{1,2}}(t;f)$ are the one-sided noise power spectral density (PSD) of the individual detectors ($\mathcal{I}_1$ or $\mathcal{I}_2$) for the data segment at time $t$. Here, $\gamma^{I}_{ft,p} $ is a geometric factor which accounts for the signal interference and the reduction in sensitivity due to the geometric nonalignment and geographic separation of the detectors, known as the overlap reduction function (ORF). It is defined as~\cite{christ92,ORF_Finn},
\begin{equation}
\gamma_{ft,p} ^{I} := \sum_{A} \int_{S^2} d \mathbf{\hat \Omega} F^{A}_{\mathcal{I}_1}(\mathbf{\hat \Omega},t) 
F^{A}_{\mathcal{I}_2}(\mathbf{\hat\Omega},t) e^{2\pi i f \frac{\mathbf{\hat \Omega}\cdot {\mathbf{\Delta x}_I (t)}}{c}} e_p(\mathbf{\hat \Omega})  ,
\label{eq_ORF_general}
\end{equation}
where $\mathbf{\Delta x}_I (t)$ is the detector separation vector. The polarizations are denoted by $A=+,\times$, and $F^{A}_{\mathcal{I}_{1,2}}(\mathbf{\hat \Omega},t)$ denotes the respective antenna pattern functions. 
Calculation of the ORF plays a crucial role in performing SGWB searches. The ORF expressed in a general basis in Eq.~(\ref{eq_ORF_general}) can be converted to the desired search basis.

\begin{figure}[ht]
\centering
\includegraphics[width=0.95\linewidth]{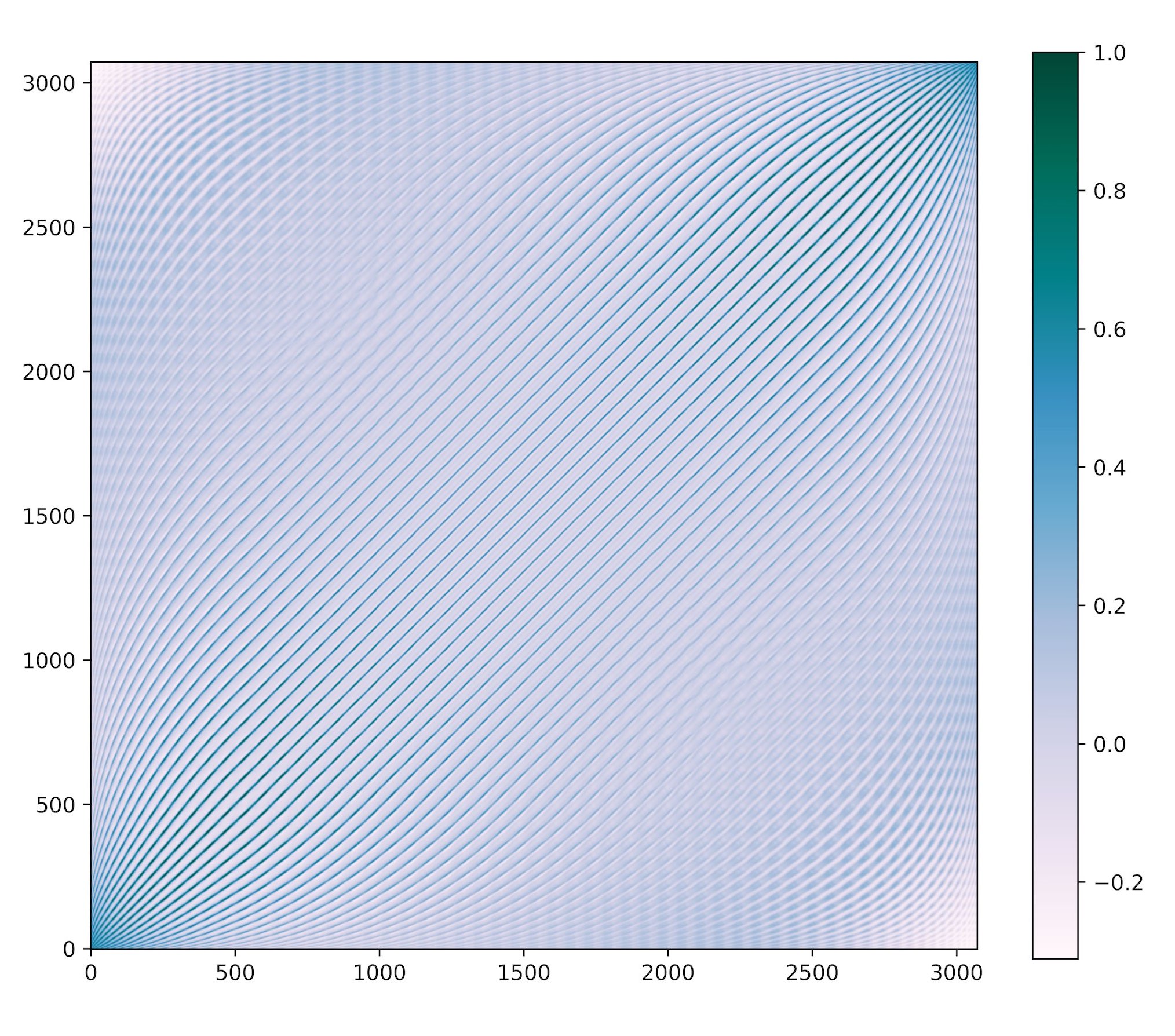}
\caption{A characteristic beam matrix for the LIGO Hanford-Livingston baseline at $\sim 3^\circ$ pixel size ($n_{\mathrm{side}} = 16$). The row and column of this square matrix represents the pixel index (3072 pixels). Each row of the matrix is the antenna response function for the pointing direction. Since the GW radiometer receives maximum contribution from the pointing direction, this matrix is dominated by the diagonal elements.}
%Here the stripes are related to the isoLatitude pixelization scheme
\label{beam_matrix}
\end{figure}

In the pixel basis, one can write the ORF as, 
\begin{equation}
\gamma_{ft,\mathbf{\hat\Omega}}^{I} \ = \ \sum_{A} F^{A}_{\mathcal{I}_1}(\mathbf{\hat \Omega},t) F^{A}_{\mathcal{I}_2}(\mathbf{\hat\Omega},t) \, e^{2\pi i f \frac{\mathbf{\hat \Omega}\cdot {\mathbf{\Delta x}_I (t)}}{c}} \,.
\label{eq_ORF_pixel}
\end{equation}
The corresponding Fisher information matrix (also called the beam matrix in the pixel basis) then becomes
\begin{equation}
\label{eq_fisher_pixel}
\Gamma_{\mathbf{\hat\Omega}, \mathbf{\hat\Omega'}} = 4 \sum_{Ift} \frac{H^2(f)}{P_{\mathcal{I}_1}(t;f) \, P_{\mathcal{I}_2}(t;f)} \,\gamma^{I*}_{ft,\mathbf{\hat\Omega}} \, \gamma^{I}_{ft,\mathbf{\hat\Omega'}} \,.
\end{equation}
The above matrix for the two LIGO detectors using a \hpx resolution $n_{\mathrm{side}} = 16$ (that divides the sky in $3072$ equal area pixels) is shown in Fig.~\ref{beam_matrix}. If the pixels are closer to the pointing direction, the beam values are stronger whereas it weakens when the distance between the pixels and pointing direction increases. This sparse nature of the matrix is reflected in the figure through the periodic stripes (the stripes in this diagonally dominated matrix can be related to the isoLatitude pixelization scheme). 
%FIGURE1
%
%FIGURE2
\begin{figure}[ht]
\includegraphics[width = 0.4\textwidth]{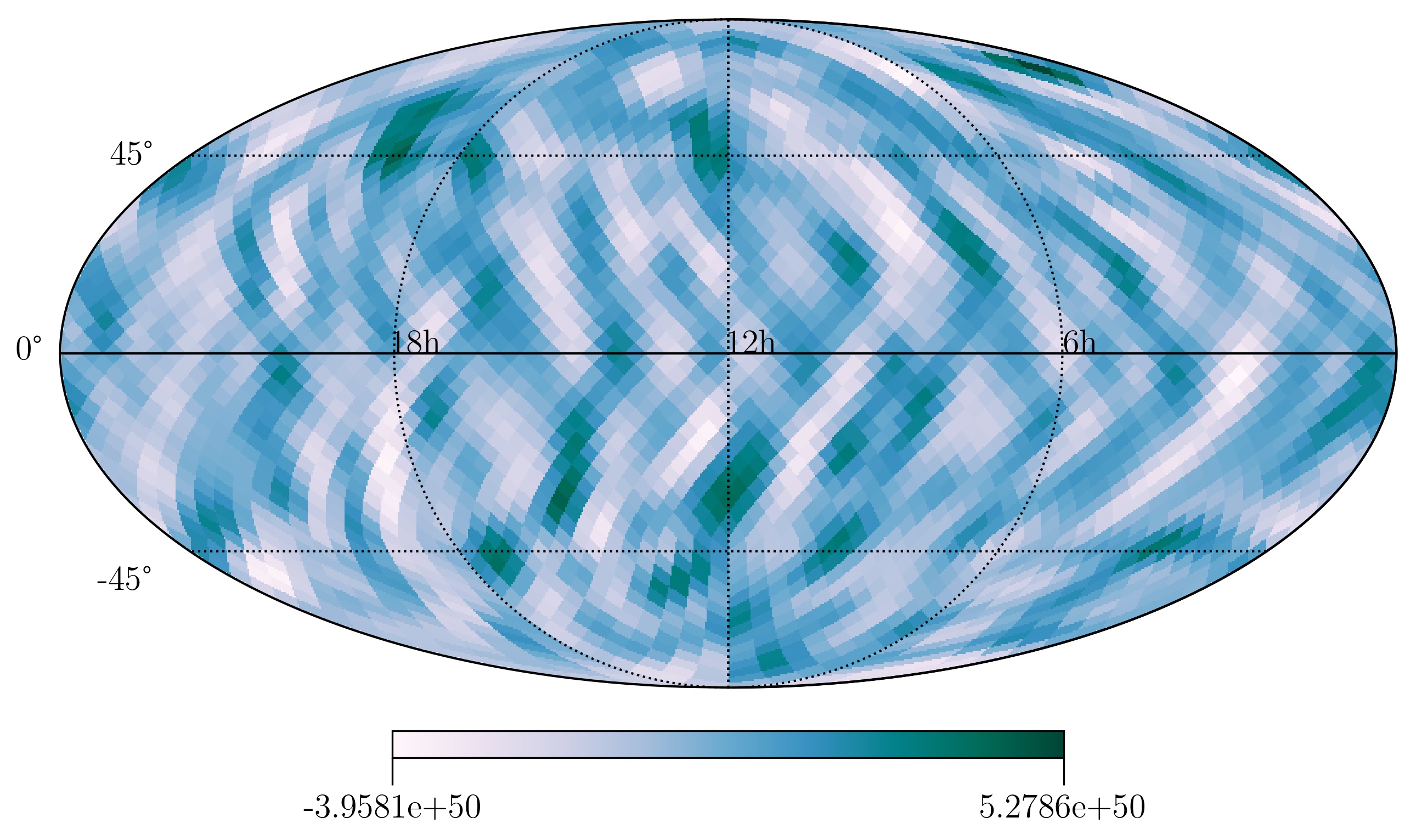}
\includegraphics[width = 0.4\textwidth]{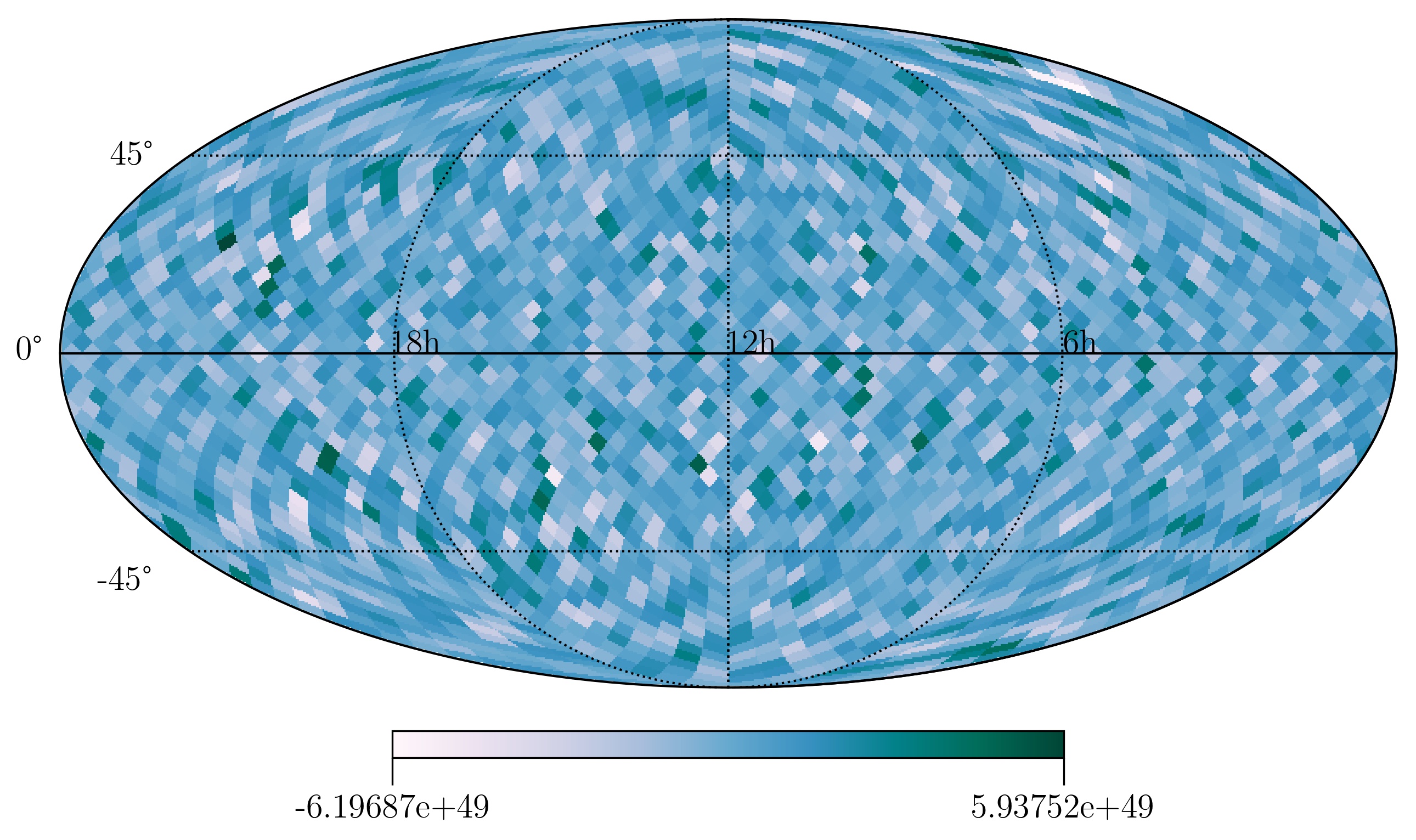}
\includegraphics[width = 0.4\textwidth]{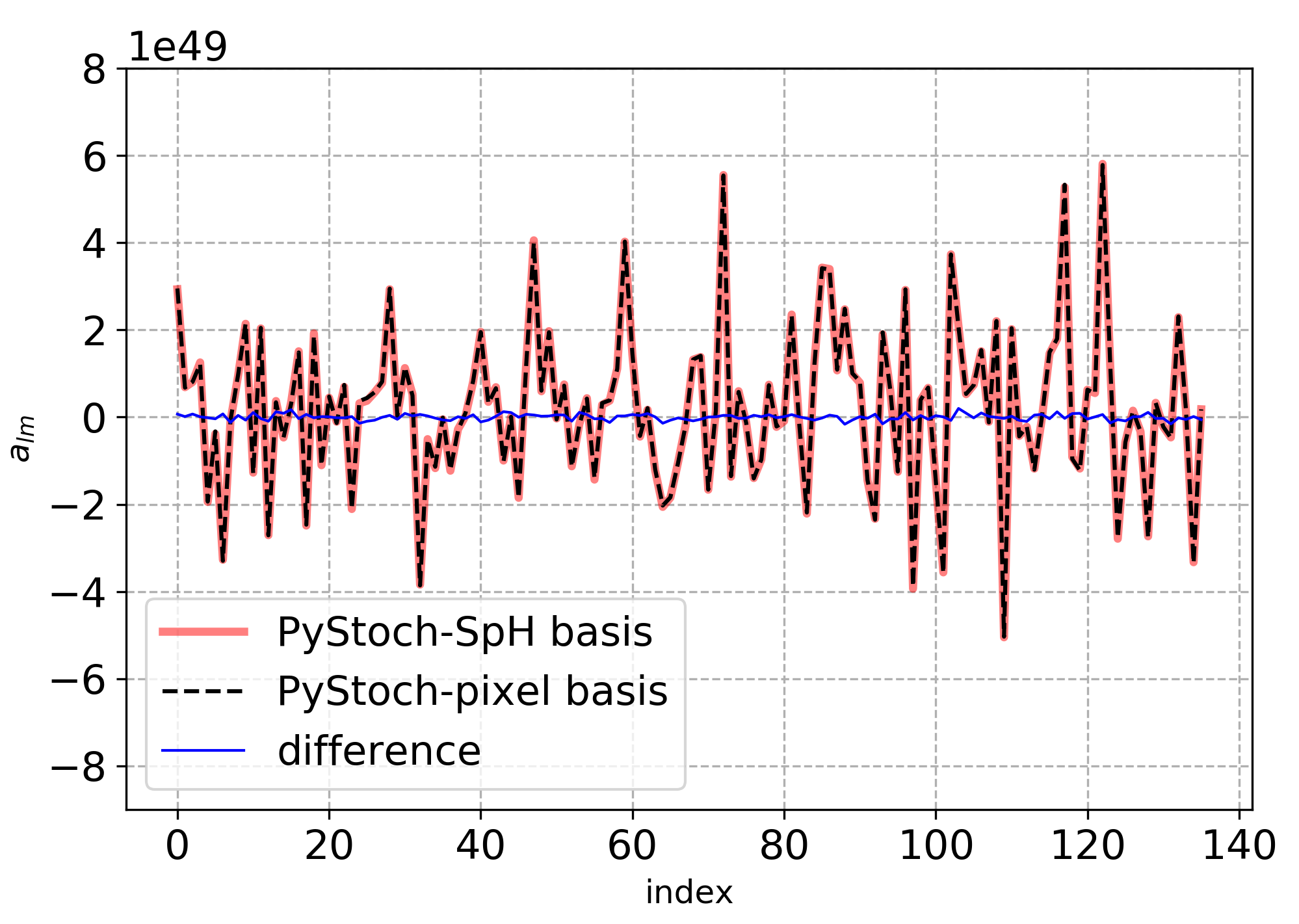}
\caption{Results from pixel to SpH transformation of the dirty map using \hpx tools are compared here. Mollweide projections of the {\tt PyStoch} pixel-space broadband map in ecliptic coordinates with $\rm{n}_{\rm{side}} = 16$ and spectral index $\alpha=3$ is shown on top and its difference from the map obtained by conventional SpH analysis in {\tt PyStoch} with $l_{\rm{max}} = 30$ in the middle. The fractional rms difference (ratio of rms of the difference map to the rms of the original map) is a few percent, as the SpH multipoles are limited to a maximum $l$ of interest (large angular scales). The main result is in the bottom panel, showing the comparison between SpH coefficients (up to $l_{\rm{max}}=15$) obtained by {\tt PyStoch} from CSD using the conventional method and from the pixel-based map, where the fractional rms differences are less than $0.1$\% (the difference between SpH coefficients estimated from CSD using {\tt PyStoch} and conventional pipeline is negligible [Fig.~\ref{alm_comparison}]). This proves that a separate sky map in SpH analysis is redundant.
}
\label{pixel_vs_sph}
\end{figure}
%
%FIGURE3
\begin{figure*}[ht]
\includegraphics[width = 0.32\textwidth]{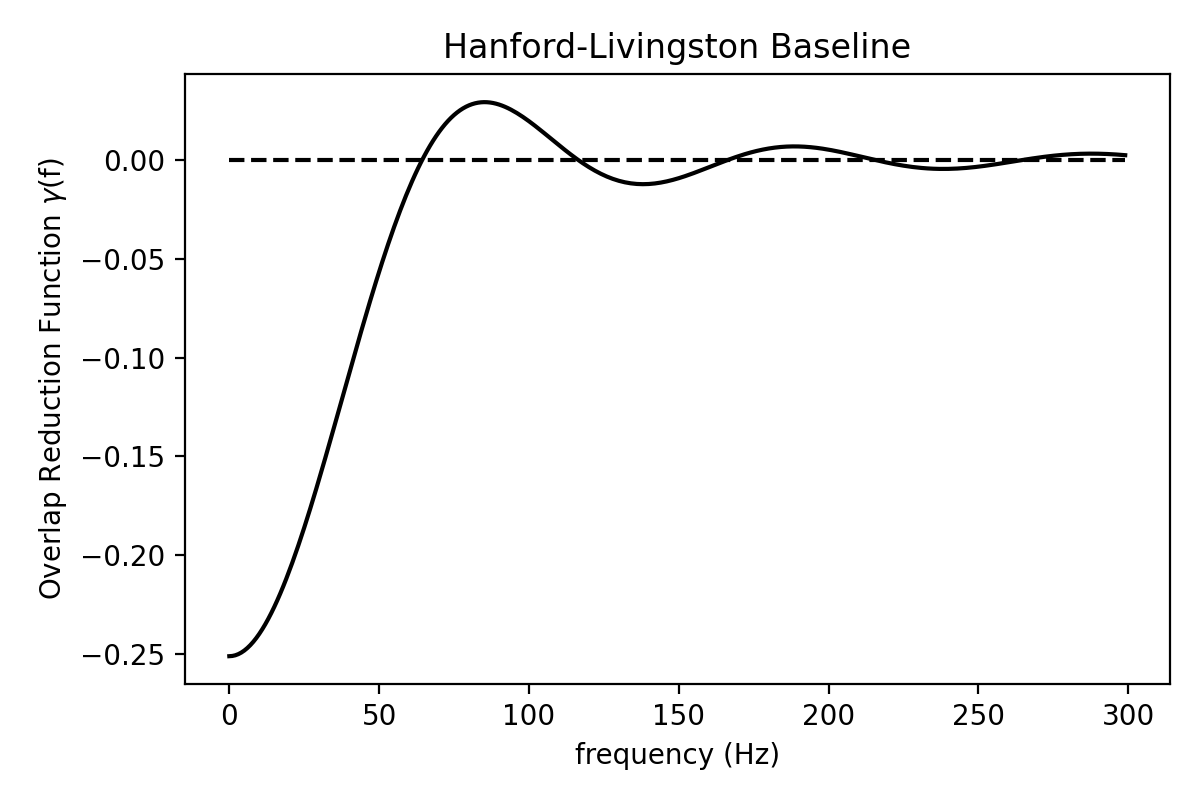}
\includegraphics[width = 0.32\textwidth]{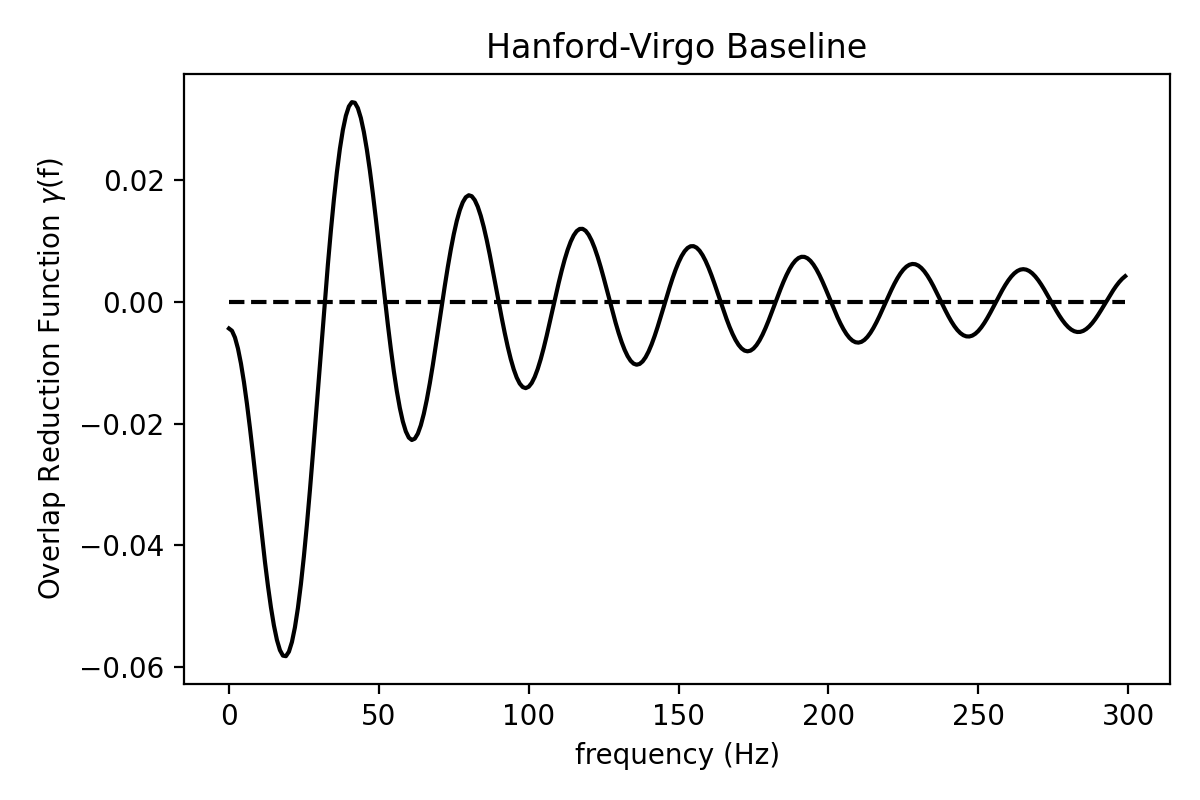}
\includegraphics[width = 0.32\textwidth]{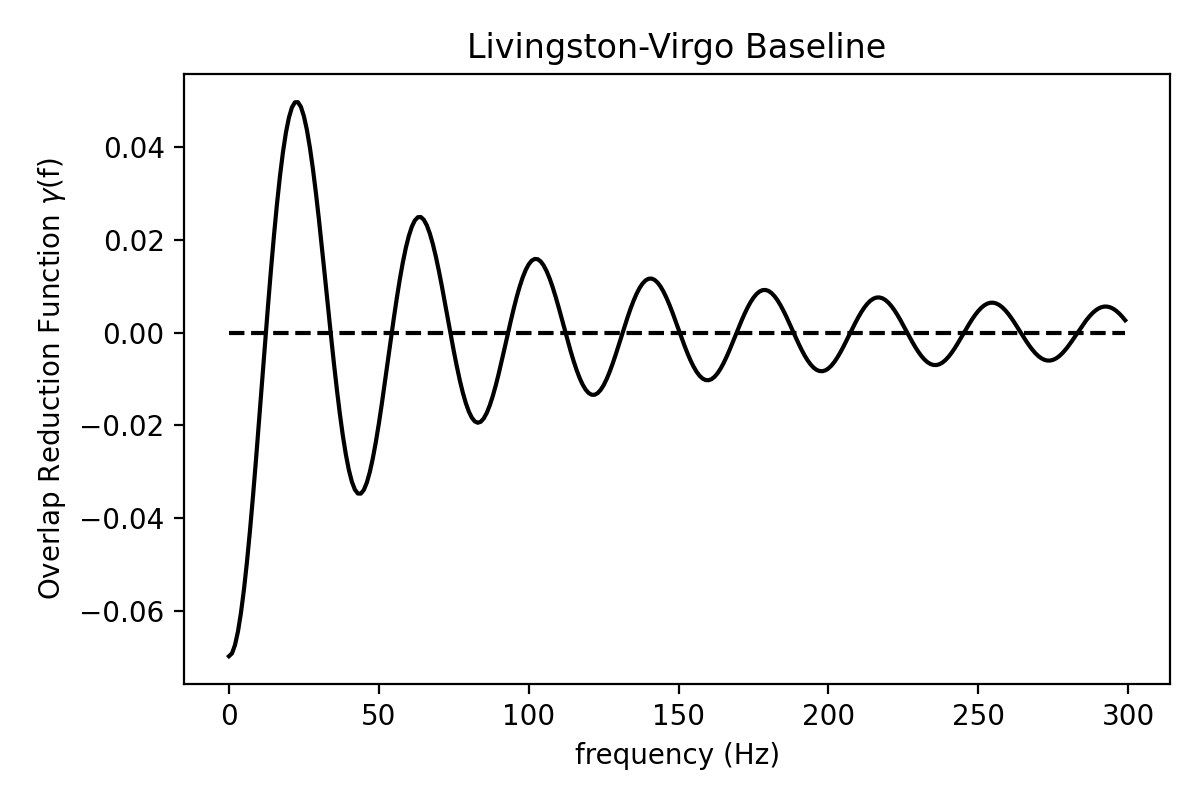} 
\caption{The first spherical harmonic term in the overlap reduction function as calculated by enhanced {\tt PyStoch} for the three possible baselines in the LIGO-Virgo detector network. The solid line is for the real part of the function, the dashed line is complex.}
\label{orf_isotropic}
\end{figure*}

In the SpH basis, expression for ORF takes the form
\begin{equation}
\gamma_{ft,lm} ^{I} = \int_{S^2} d \mathbf{\hat \Omega} \, \gamma_{ft,\mathbf{\hat\Omega}}^{I} \, Y^{*} _{lm}(\mathbf{\hat\Omega}) \,.
\label{eq_ORF_spherical}
\end{equation}
and the Fisher information matrix elements are given by,
\begin{equation}
\label{eq_fisher_sph}
\Gamma_{lm,l'm'} = 4 \sum_{Ift} \frac{H^2(f)}{P_{\mathcal{I}_1}(t;f) \, P_{\mathcal{I}_2}(t;f)} \,\gamma^{I*}_{ft,lm} \, \gamma^{I}_{ft,l'm'} \, .
\end{equation}
For odd values of $l+l'$, these elements vanish~\cite{Thrane09}. One can use $\gamma_{ft,lm}^{I}$ obtained below to compute the nonzero elements of the matrix.

It is, therefore, possible to transform the results in the pixel basis to the SpH basis using the following formulas:
\begin{eqnarray}
    X_{lm} &=& \int_{S^2} d \mathbf{\hat \Omega} \, Y^{*} _{lm}(\mathbf{\hat\Omega}) \, \gamma_{ft,\mathbf{\hat\Omega}}^{I} \,,\\
    \Gamma_{lm,l'm'} &=& \int_{S^2} d \mathbf{\hat \Omega} \int_{S^2} d \mathbf{\hat \Omega'} \, Y^{*} _{l'm'}(\mathbf{\hat\Omega}') Y_{lm}(\mathbf{\hat\Omega}) \, \Gamma_{\mathbf{\hat\Omega}, \mathbf{\hat\Omega'}} \, . \ \
\end{eqnarray}
For this work, we first considered transforming the pixel-to-pixel Fisher matrix to the SpH domain. However, this seemed more computationally challenging than using SpH transform of the pixel-based overlap reduction functions [Eq.~(\ref{eq_ORF_pixel})], which are already computed for the pixel-based search, and could also be used to obtain the estimates of the SpH moments of the sky directly.

Since the change in the ORF in SpH basis due to the rotation of Earth in time $t$ is equivalent to increasing the azimuthal angle $\phi$ about the spin axis by $- 2\pi t/T$, following the definition of spherical harmonics, one can write~\cite{AllenOttewill,Romano2017},
\begin{equation}
\gamma_{ft,lm}^{I} = \gamma_{f0,lm}^{I} \, e^{im2\pi t/T} \,,
\label{eq_ORF_time_dependence_gnrl}
\end{equation}
where $T$ is the period of rotation of Earth, which is by definition one sidereal day.
Thus $\gamma_{ft,lm}^{I}$ can be computed at all values of $t$, using $\gamma_{f0,lm}^{I}$, the spherical harmonic transform of $\gamma_{ft,\mathbf{\hat\Omega}}^{I}$ at a fiducial $t=0$. This step significantly reduces computation cost, by alleviating the need to perform a spherical harmonic transform of $\gamma_{ft,\mathbf{\hat\Omega}}^{I}$ at each of the $\sim 1000$ segments in one sidereal day. This is particularly important for the new all-sky-all-frequency (ASAF) search, where this quantity must be computed at every frequency bin, which are a few tens of thousands in number for the frequency bin size of $1/32$Hz presently being used in LIGO-Virgo-KAGRA analyses. With the help of this technique, we have enabled {\tt PyStoch} to perform ASAF search in the SpH basis, along with the pixel basis (which it was already capable of doing). While the SpH moments estimated through this route match those obtained directly from the pixel-based map (Fig.~\ref{pixel_vs_sph}), these tools will nevertheless be required for estimating statistical quantities in the SpH basis and will remain useful for faster execution of (perhaps exploratory) analyses where the pixel-based maps may not be necessary.

Note that our implementation is different from the existing method~\cite{Thrane09} which uses a semianalytical approach; however, the results match to better than one part in $10^4$ in terms of root mean square (rms) difference between the SpH moments and the Fisher matrices (Figs.~\ref{pystoch_sph_dirtymap} and \ref{pystoch_sph_fisher}). Our method, powered by fast-Fourier transform employed by \hpx, is highly efficient [details in Sec.~\ref{calc_ORF}]. However, due to the lack of a straightforward recipe, we have not compared the relative performances of the two approaches.

%---------------------------------------------------------------------
\section{Implementation}
\label{sec:implementation}

We have added the new methods for estimating SpH moments in {\tt PyStoch}. The first method, to get SpH moments from the pixel-based dirty map, is trivial using \hpx tools. The second method is more involved but useful for various reasons as mentioned above (primarily because transforming the pixel-to-pixel Fisher information matrix to the SpH basis may be numerically challenging). With the new tools, {\tt PyStoch} can efficiently compute results in both SpH and pixel bases together in less than two hours on a single CPU (with multiple cores) for the whole observation run data folded to one sidereal day at the usual angular resolution, including the full Fisher information matrix in the SpH basis.\footnote{Computation cost for pixel-to-pixel Fisher information matrix is also small in {\tt PyStoch}, but parallel processing has not been implemented yet for this part of the code due to memory issues. If this item is included, it takes a few hours on a single CPU core to compute all the results (maps, SpH moments, and Fisher matrices).}

The results have been validated using the GW data from the first observing run of LIGO Hanford-Livingston detectors. We have created the mentioned dataset by following the steps described in~\citet{Ain_Folding}. 
The dataset consist of $898$~frames of CSDs and PSDs, each with a $50$\% overlapping segment duration of $192$~sec, spanning nearly one full sidereal day. Considering the optimal resolution required for the radiometer analysis for the two LIGO detectors, we choose the \hpx map resolution to be $n_{\mbox{side}} = 16$ (for illustration purposes we set $n_{\mbox{side}} = 32$ in all the sky maps). This corresponds to $n_{\rm pix} = 3072$ pixels for the entire sky; each pixel is nearly square with a width of approximately $3$ degrees.

For a source whose intensity varies slowly across the sky, $l_{\rm{max}}$ could be small; on the other hand, for sources with sharply varying spatial distributions, $l_{\rm{max}}$ is expected to be large. A recommended relation between pixel and SpH resolution is~\cite{HEALPix}
 \begin{equation}
 \label{eq_lmax_reco}
    n_{\rm pix} = 12\, n_{\mbox{side}}^2  \approx \frac{4}{\pi} \, l _{\rm{max}}^2 \,.
 \end{equation}
Although the equivalent $l_{\rm{max}}$ corresponding to $n_{\mbox{side}} = 16$ is higher according to Eq.~(\ref{eq_lmax_reco}), we have used an $l_{\rm{max}} = 30$ for our study (here, $l_{\rm{max}}$ is chosen somewhat arbitrarily, and the choice is consistent with previous LIGO-Virgo Collaboration stochastic studies~\cite{O1directional,O2directional}).

The excellent match between SpH coefficients estimated directly from CSD and from the pixel-based dirty map is shown in Fig.~\ref{pixel_vs_sph}, firmly establishing that a separate sky map for SpH basis analysis is redundant.

%
%-----------------------------------------------------------------------
\subsection{Calculating Overlap Reduction Function}
\label{calc_ORF}
{\tt PyStoch} uses a novel technique for calculating the ORF through seed matrices~\cite{pystoch}, since the whole ORF with three independent indices (direction, time, and frequency) would constitute a very large array that cannot be saved to disk or read from it efficiently. The time-dependent part of the ORF in Eq.~(\ref{eq_ORF_general}) can be separated from the frequency-dependent part by calculating $F^{A}_{\mathcal{I}_1}(\mathbf{\hat \Omega},t) 
F^{A}_{\mathcal{I}_2}(\mathbf{\hat\Omega},t)$ and
${\mathbf{\hat \Omega}\cdot {\mathbf{\Delta x}_I (t)}}/{c}$ separately as maps in the basis we are using. For a fixed time segment and baseline, these two maps are frequency independent. We call these maps ORF seed matrices. {\tt PyStoch} calculates and saves them for each set of data and automatically loads them for subsequent analysis. 
The actual ORF can be quickly calculated in a loop over frequencies from the seed matrices. This method of calculating the ORF is perhaps the fastest with reasonable memory usage. 

To calculate the ORF in the SpH basis we use the existing {\tt PyStoch} module of calculating the ORF in the pixel basis and then converting it in the SpH basis using \hpx tools. This method is extremely fast and accurate and does not require dealing with SpH formulas (\hpx tools do that internally).
However, the regular \hpx tools for calculating SpH maps from pixel maps work for real maps only, so we had to wrap that routine into a new one to handle complex ORF maps.

After performing the basis conversion of the ORF for the first time segment, calculating the ORF for other time segments can be made faster by exploiting the Earth rotation time dependency given in Eq.~(\ref{eq_ORF_time_dependence_gnrl}). This way, by using ORF seed matrices and azimuthal symmetry, the ORF in the SpH basis can be calculated very accurately and very quickly even for very high $l_{\rm{max}}$ (see Fig.~\ref{orf_isotropic}). This unification of techniques from the two methods enables us to efficiently create maps in both pixel and SpH bases.

\begin{figure}[h]
    \centering
    \includegraphics[width=0.45\textwidth]{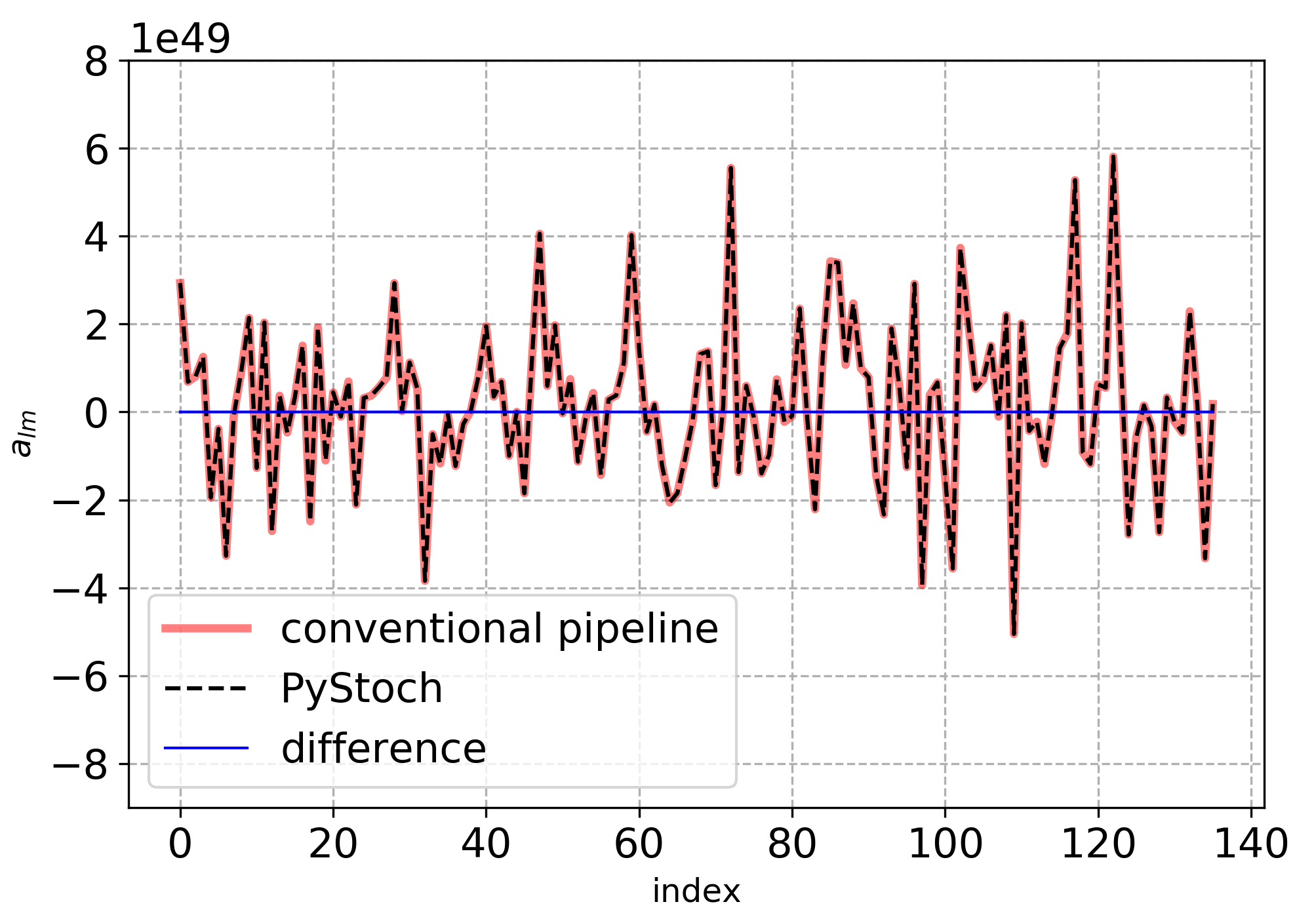}
    \caption{Comparison of SpH coefficients $a_{lm}$, computed from two different pipelines (in contrast to Fig.~\ref{pixel_vs_sph}, where the comparison is with SpH coefficients from the pixel-based dirty map). Here, we restrict the results up to $l_{\rm{max}}=15$, corresponding to $256$ total modes. The thick solid line represents $a_{lm}$ computed using the conventional pipeline whereas the black dashed line is obtained using {\tt PyStoch}. The differences in their values, shown with a thin solid line, are nearly zero. An rms difference of a few times $10^{-5}$ (as was the case for dirty maps in Fig.~\ref{pystoch_sph_dirtymap}) validates the SpH coefficients, as well as the dirty map computed using the new pipeline. }
    \label{alm_comparison}
\end{figure}
\begin{figure}[hb]
    \centering
    \includegraphics[width = 0.4\textwidth]{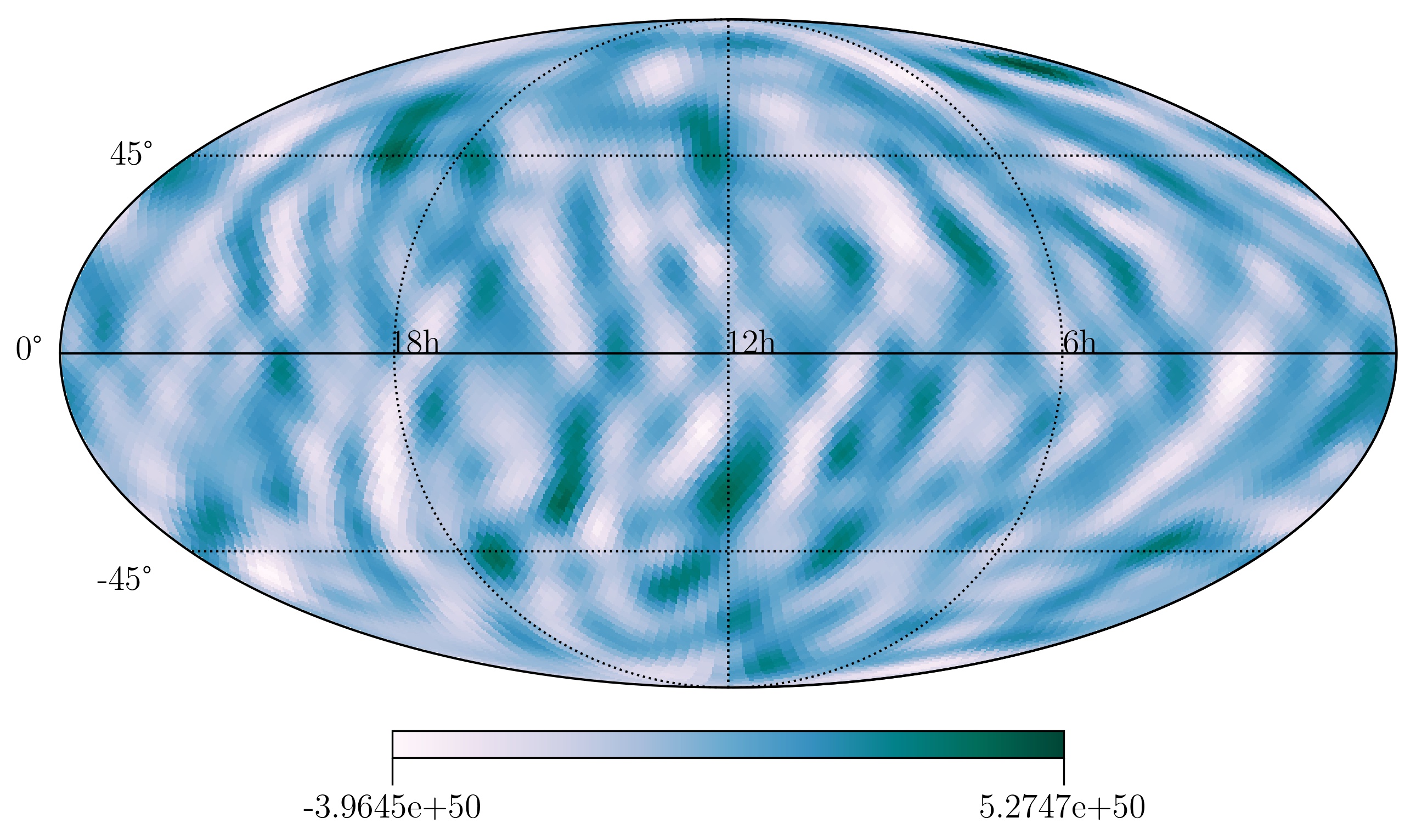}
    \includegraphics[width = 0.4\textwidth]{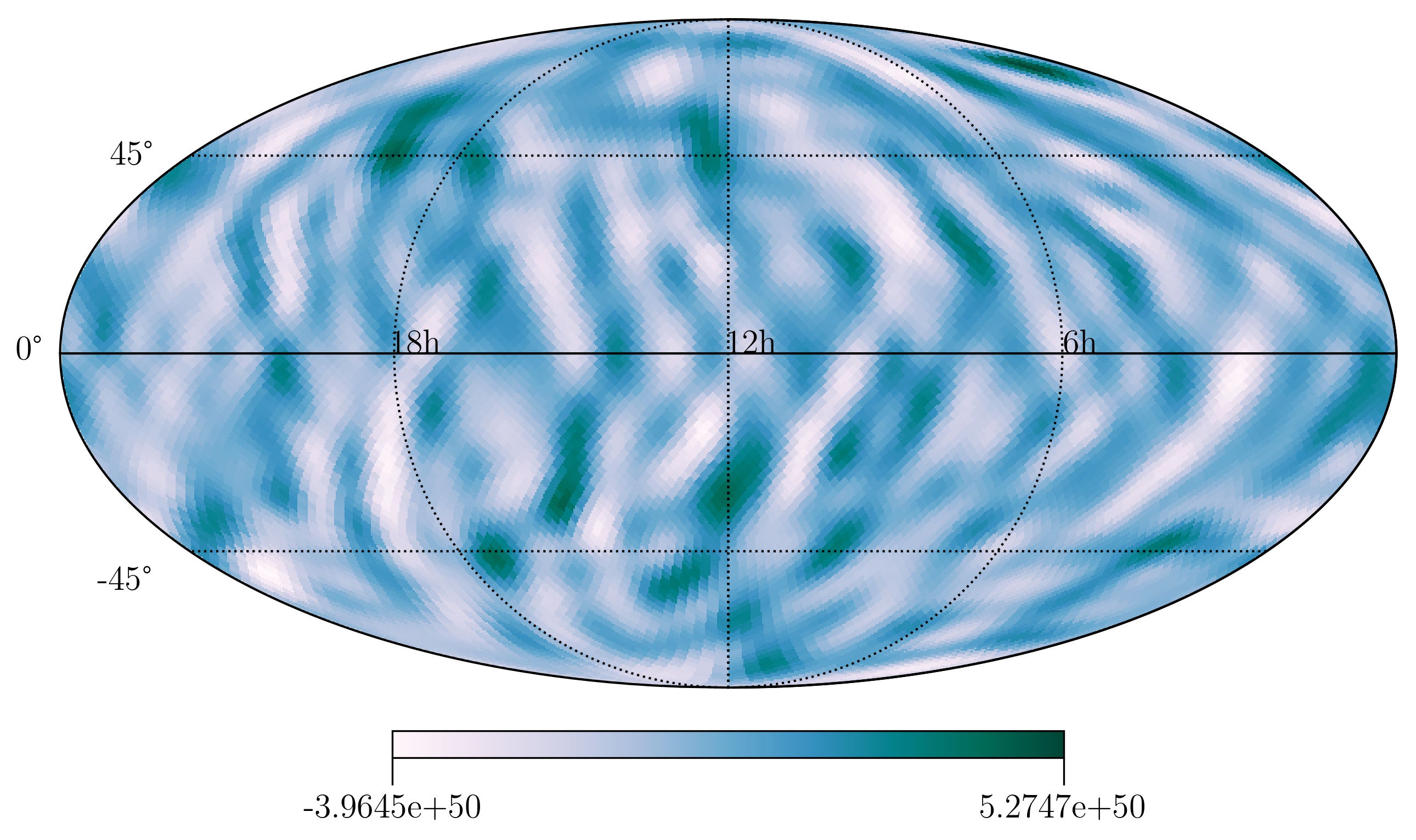}
    \includegraphics[width = 0.4\textwidth]{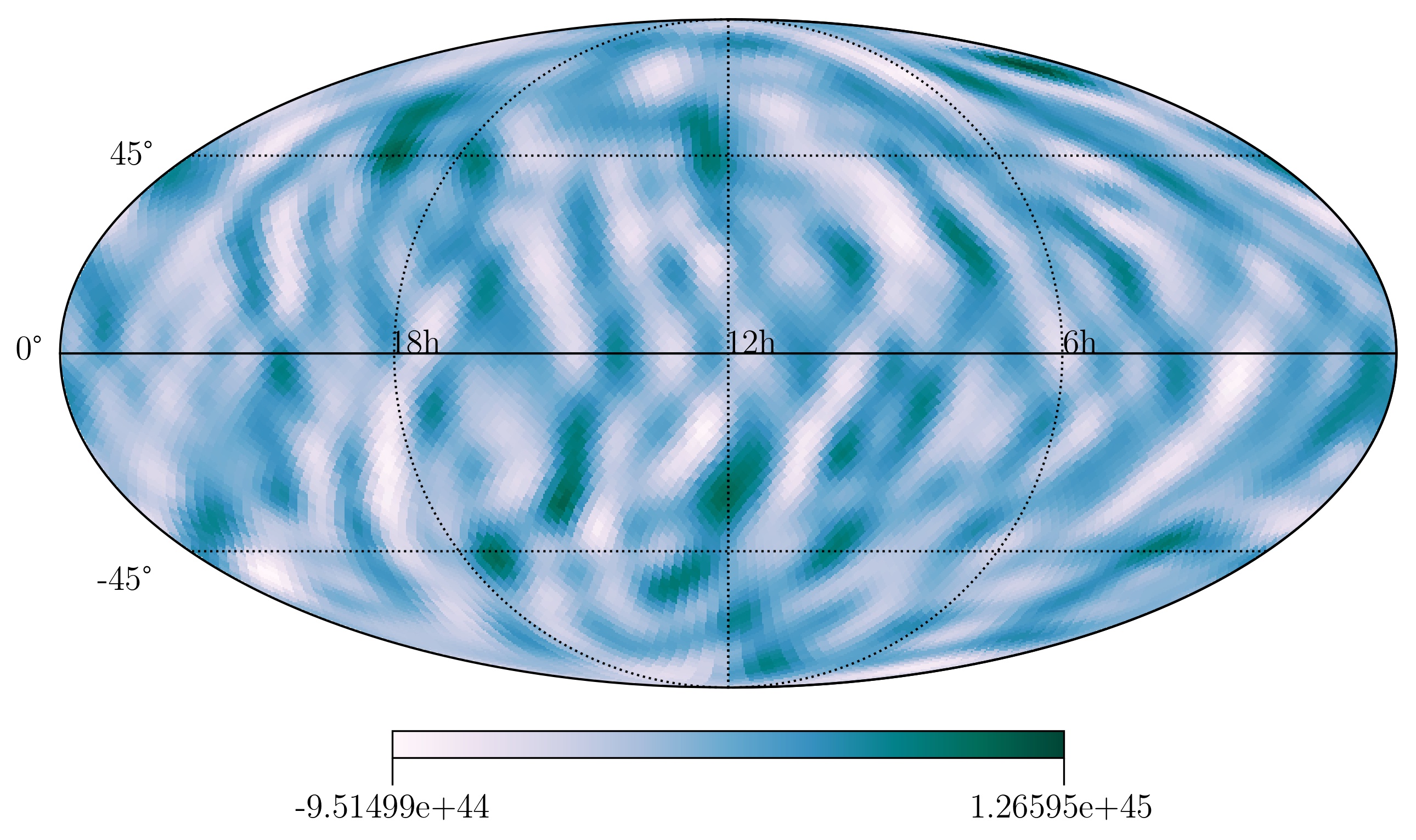}
    \caption{The dirty maps produced from SpH coefficients using the conventional pipeline is shown on the top whereas the one from  the {\tt PyStoch} pipeline is at the middle. On the bottom, we show the difference between the maps. The rms difference of these maps is of the order of a few times $10^{-5}$. All the maps are represented as a color bar plot on a Mollweide projection of the sky in ecliptic coordinates with $\rm{n}_{\rm{side}} = 32$, power-law spectral indices $\alpha=3$ and have $l_{\rm{max}} = 30$.}
    \label{pystoch_sph_dirtymap}
\end{figure}

%
%
%----------------------------------------------------------------------------
\subsection{Calculating the dirty map}
Starting from Eq.~(\ref{eq_dirtymap}), the expression of the dirty map can be simplified as,
\begin{equation}
X_p = \sum_{If\tsid} K^{I*}_{f\tsid,p} \, x^I_{f\tsid} \,, \label{fwX}\\
\end{equation}
where $K^{I}_{ft,p} \ := \ \tau \, H(f) \, \gamma^{I}_{ft,p}$ is the mapping kernel and $x^I_{f\tsid}$ is folded time frequency data~\cite{Ain_Folding}, which are calculated from the CSD $\widetilde{s}_{\mathcal{I}_1}^*(t;f) \widetilde{s}_{\mathcal{I}_2}(t;f)$ and $\tau$ represents the entire observation duration. This compression is explained in detail in Ref.~\cite{Ain_Folding}.
The calculation of the dirty map is straightforward once the ORF has been calculated. Just by changing the basis of the ORF from pixel to SpH, the basis of the resulting dirty map changes without requiring any modification in the {\tt PyStoch} algorithm.

Previously, we noted and demonstrated that calculation of SGWB sky maps can be done in a few minutes on an ordinary laptop~\cite{pystoch}. The SpH search we present here is not only an order of magnitude faster than the conventional SpH search owing to folded data and the way it is handled by {\tt PyStoch}, it is even faster than the {\tt PyStoch} pixel search because the SpH maps have fewer terms compared to the pixel maps (961 for $l_{\rm{max}}=30$ compared to 3072 pixels).

In order to make sure that the enhanced {\tt PyStoch} pipeline with unified mapmaking capability produces similar results as the one used in Refs.~\cite{O1directional,O2directional}, we did some investigation on the dirty maps obtained from the new pipeline and the conventional one~\cite{gitpublic}. In Fig.~\ref{pystoch_sph_dirtymap}, we show the dirty maps obtained from the two pipelines. To further demonstrate the validity of the upgraded {\tt PyStoch}, we compared the spherical harmonic coefficients obtained from these two different methods in Fig.~\ref{alm_comparison}. It is evident that the obtained results are identical, validating the dirty mapmaking part of our pipeline.

%%
%----------------------------------------------------------------------------
%
\subsection{Calculating the Fisher information matrix}

\begin{figure*}
\includegraphics[width = 0.325\textwidth]{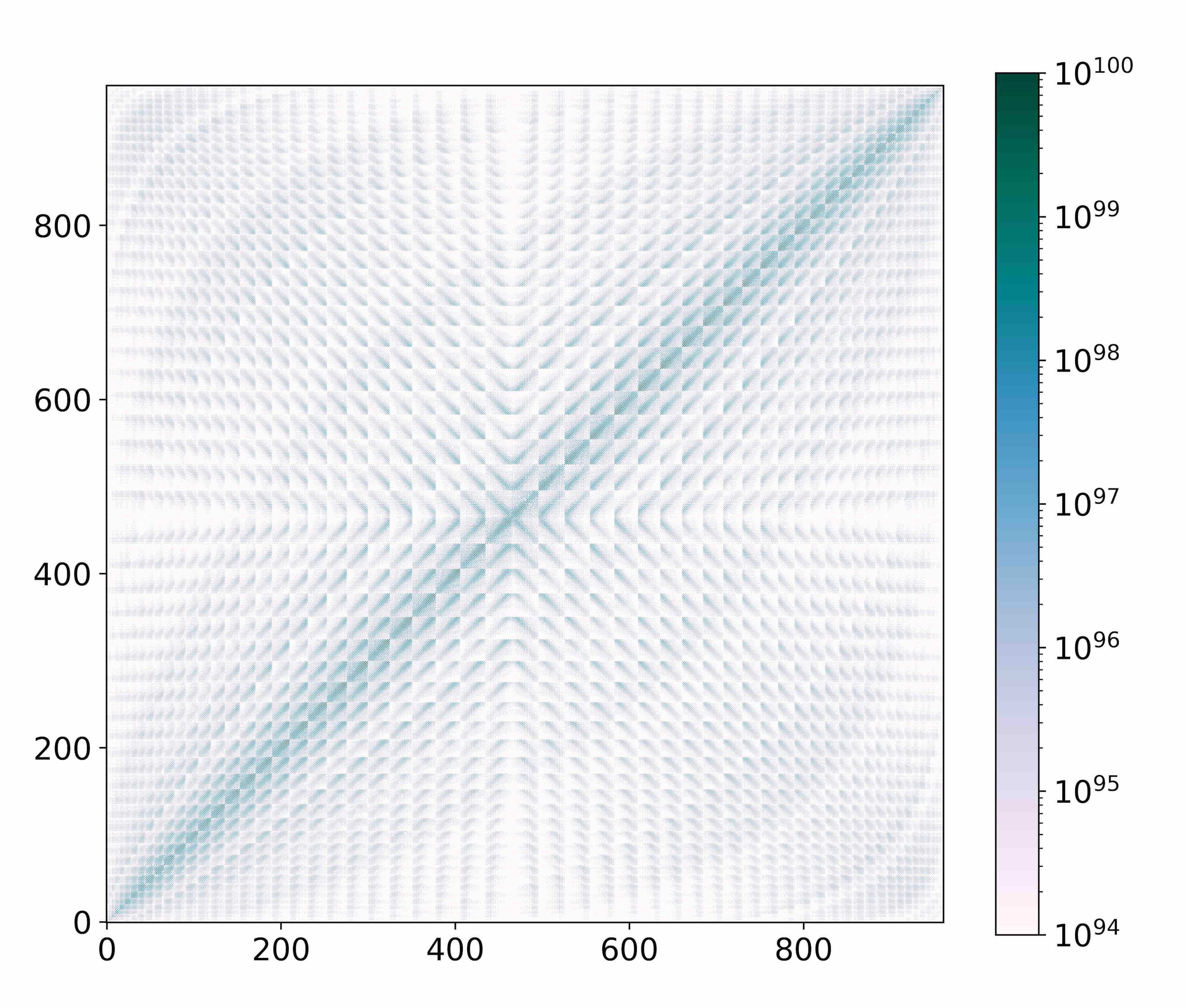}
\includegraphics[width = 0.325\textwidth]{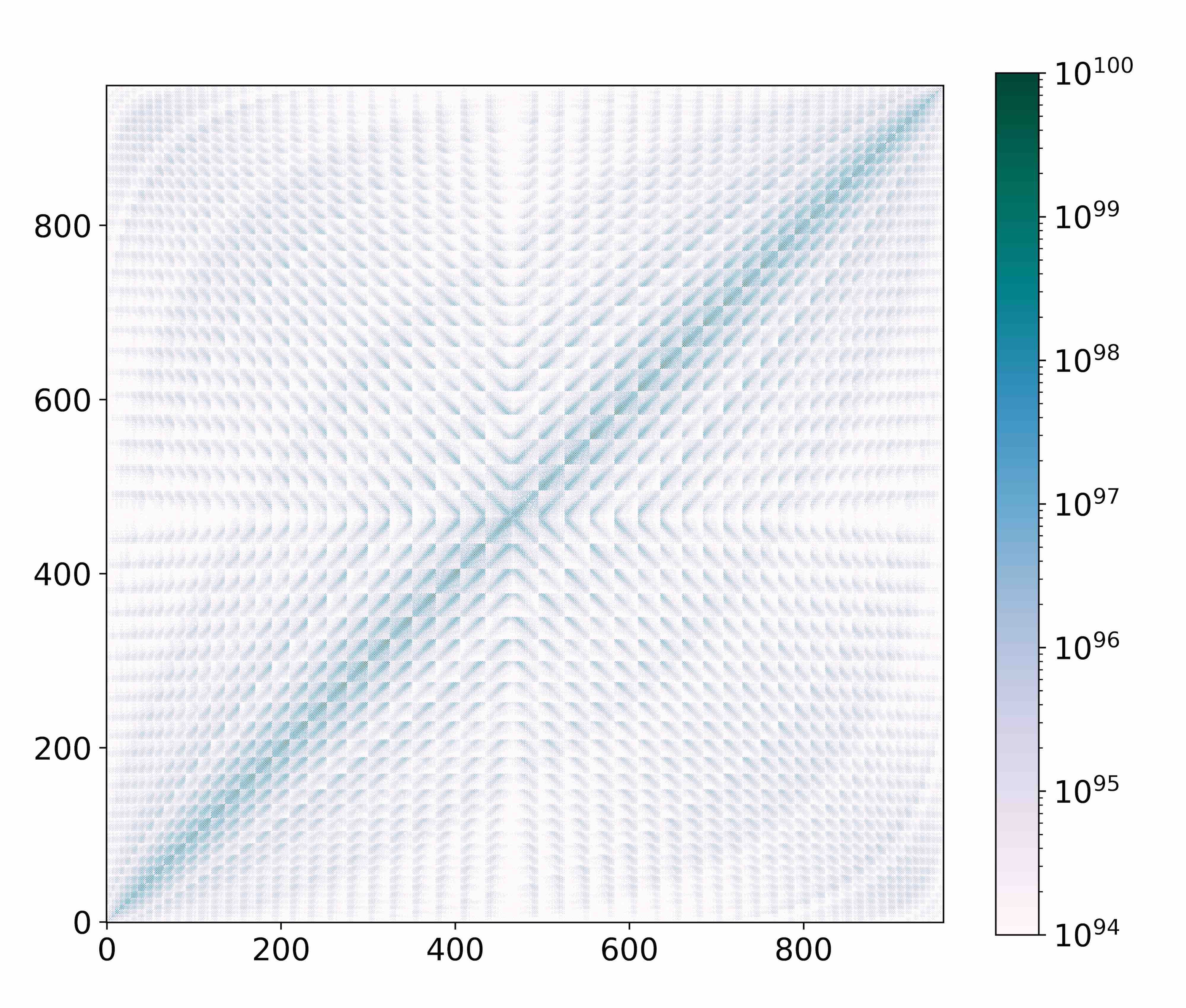}
\includegraphics[width = 0.325\textwidth]{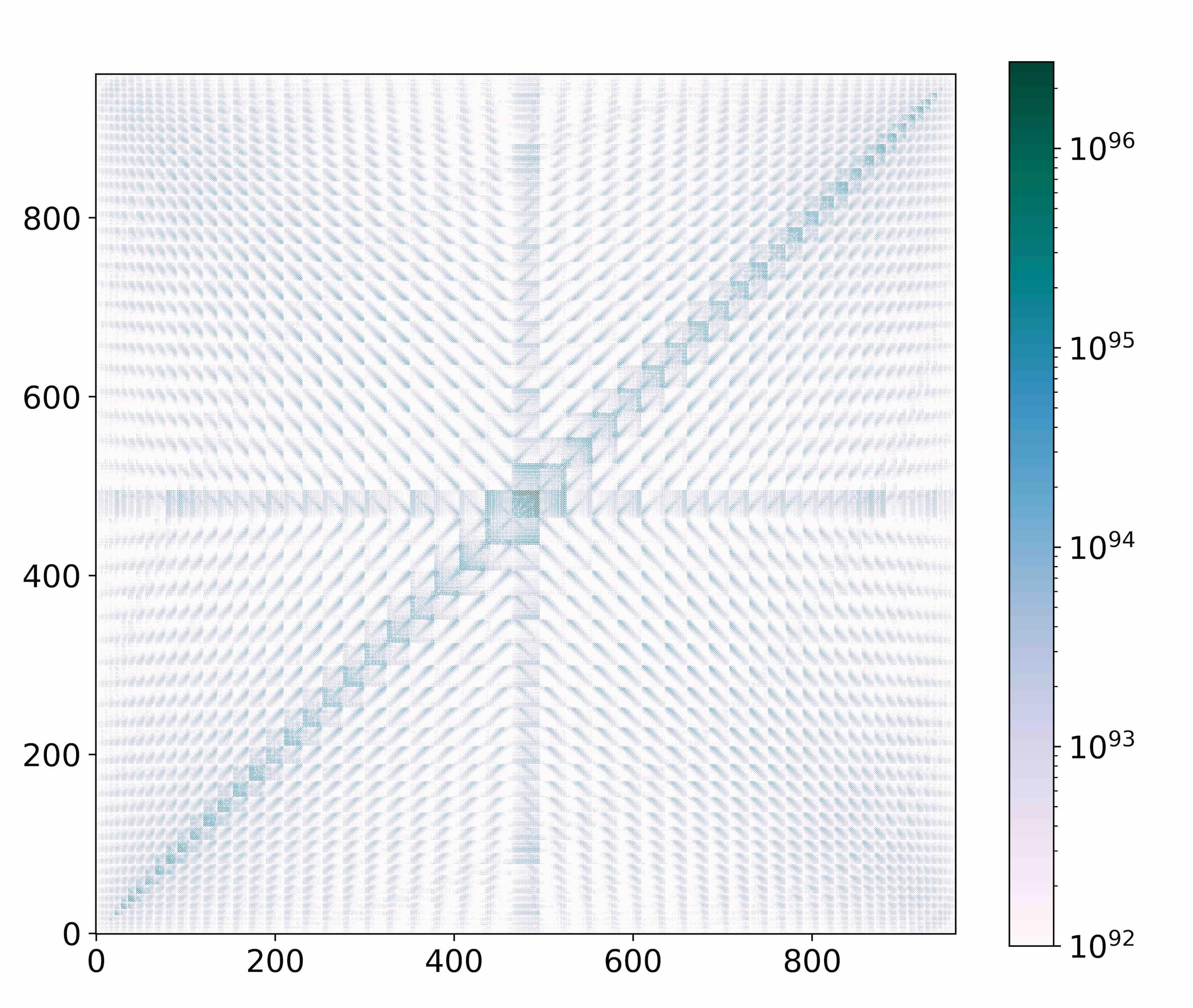}
\caption{Comparison of the Fisher matrix, with both the axes denoting the index (the indexing follows the same convention used in ~\citet{Thrane09}) corresponding to the observed mode [for this analysis, $l_{\mbox{max}}=30$ correspond to $(l_{\mbox{max}}+1)=961$ modes], from two pipelines is shown in this figure. The Fisher matrix (absolute value) produced using the conventional pipeline is shown on the left, whereas the one from {\tt PyStoch} is in the middle. On the right, we show the absolute difference between the Fisher matrices, whose small values validate the consistency between the pipelines (elements with too small absolute differences have been masked to avoid the divergence of the $\log$ of a value close to zero). From the rms difference of these matrices, the agreement can be quantified to be better than one part in $10^4$
for both real and imaginary parts of the Fisher matrices.
Here, the power-law spectral index and $l_{\rm{max}}$ take the same values as of the dirty maps in Fig.~\ref{pystoch_sph_dirtymap}.}
\label{pystoch_sph_fisher}
\end{figure*}

The Fisher information matrix is necessary for estimating the error on the sky map and the spherical harmonic moments and also to produce deconvolved ``clean'' maps.
The calculation of the Fisher information matrix (or the beam matrix in the pixel basis) can be understood as the equivalent of making one dirty map for each pixel by placing a unit point source at that pixel. Even though the computational cost of this calculation scales as the square of the number of pixels, using the algebraic method proposed in Ref.~\cite{pystoch} we were able to calculate them for our analysis. The challenge to obtain the Fisher matrix in the SpH basis becomes trivial, given we have now calculated the ORF and dirty maps in the SpH basis. Following the recipe from Sec.~\ref{sec:method}, we have successfully calculated this matrix.
In Fig.~\ref{pystoch_sph_fisher}, we present the Fisher information matrix calculated using the conventional pipeline and {\tt PyStoch}. From this figure, it is evident the results are essentially the same. One can now use this Fisher matrix to obtain the clean map, study the corresponding angular scale of the structure found in the map, and put upper limits.

The Fisher matrix for a single baseline of two detectors has poorly observed modes, which makes direct inversion of the Fisher matrix (Fig.~\ref{beam_matrix} \& \ref{pystoch_sph_fisher}), impractical. ML estimation of the true SGWB sky takes the simple form given in Eq.~(\ref{eq_ML_solution}), only when the inverse of the beam matrix exists. To obtain the true SGWB sky map, one has to deconvolve the dirty maps either by linearly solving the convolution equation~\cite{Mitra07} or by applying appropriate regularization in $\mathbf{\Gamma}$ before the inversion~\cite{regDeconv}. {\tt PyStoch} is capable of handling any standard regularization [norm-regularisation, gradient regularisation, Singular Value Decomposition (SVD), etc.] to condition the Fisher matrix $\Gamma'_{lm,l'm'}$, though comparing the relative performances of these deconvolution schemes in the analysis of present data is outside the scope of this paper. For completeness, we have illustrated this procedure for SVD regularization case in the Appendix. 

%%
%%
%----------------------------------------------------------------------------
\section{Conclusions}
\label{sec:conclusion}

Constraints on anisotropic stochastic backgrounds are routinely put using data from ground-based gravitational wave observatories by estimating maps and spherical harmonic moments of the sky using cross-correlation-based algorithms. In these analyses, two sets of maps used to be published, one from pixel domain analysis and one from spherical harmonic domain analysis. Here we first establish that SpH moments of the sky map and direct estimate of the SpH moments from cross-spectral density data using conventional analysis match very well. Moreover, using a modified algorithm, we show that the direct estimate of SpH moments can be efficiently and precisely obtained utilizing some of the computationally intensive quantities that are already generated for pixel domain mapmaking. We incorporate a proper basis transformation using the mathematical symmetries identified in past formalisms. We validated that this basis transformation works accurately for the maps and the Fisher information matrices. 
The derived upper-limit maps would match within $\sim 0.1\%$, while even a few percent deviation would be generally acceptable.
We introduce both the methods for estimating SpH moments of the SGWB sky in {\tt PyStoch} to enhance its capabilities. {\tt PyStoch} is being used by the LIGO-Virgo-KAGRA Collaboration for making sky maps from the latest datasets~\cite{O3-BBR}.

Apart from having the convenience of being able to produce both the measures of anisotropy, maps, and SpH moments, together, with enormous computational efficiency, this unified scheme introduces a better scope for comparing and validating the results while maintaining statistical accuracy. From now on, multiple SGWB maps need not be produced from the data. The tools we incorporate in {\tt PyStoch} will be sufficient to perform the standard SpH analysis on the latest dataset. Furthermore, now the SpH-based search will also be able to take advantage of powerful \hpx tools, commonly used in cosmic microwave background analyses.

Even though in this paper we have restricted to one spectral shape of the modeled power spectral density of the source, it has been shown previously that the analysis using {\tt PyStoch} mitigates the need to perform separate searches for different spectral shapes~\cite{pystoch}. Since these properties are preserved in the enhanced {\tt PyStoch}, one can trivially perform a model-independent mapping of the SGWB sky also in the SpH basis. 

The implementation of the SpH search in {\tt PyStoch} has unified and streamlined the different kinds of stochastic searches in an unprecedented way. It is now ready to perform any 2-sphere analysis to search for SGWB anisotropy (narrow band, or broadband) the literature has to offer. Another example of how multiple searches have been united would be to consider the isotropic search. One needs only to run {\tt PyStoch} in the SpH basis with $l_{\rm{max}}=0$. We can now get the SpH and isotropic search results at every frequency bin as well as for the broadband analyses. SGWB results, although not impressive in spatial resolution, are complex in structure. It has incredible cosmological and astrophysical information hidden in it and may also contain signatures of unknown persistent sources. A lot of studies, e.g., lensing, multipoint correlation, and polarization remain to be performed on them. Our effort of unifying the preliminary analysis of SGWB provides a robust launch pad for future studies.
%
%
%----------------------------------------------------------------------------
\begin{acknowledgments}
The authors thank Joe Romano for carefully reading the manuscript and providing valuable comments. This work significantly benefited from the interactions with the Stochastic Working Group of the LIGO-Virgo-KAGRA Scientific Collaboration. We acknowledge the use of IUCAA LDAS cluster Sarathi for the computational and numerical work. J.S. acknowledges the support by JSPS KAKENHI Grant Number JP17H06361 and expresses thanks to Hideyuki Tagoshi and Hirotaka Yuzurihara for the helpful discussion. A.A. acknowledges support by INFN Pisa and European Gravitational Observatory (EGO) and thanks Giancarlo Cella for his support. S.M. acknowledges support from the Department of Science and Technology (DST), India, provided under the Swarna Jayanti Fellowships scheme. This article has a LIGO document number LIGO-P2000461.
\end{acknowledgments}
%----------------------------------------------------------------------------
%
\appendix
\section{SVD regularization and $C_l$ estimates}
\label{sec:SVD}

\begin{figure}
    \centering
    \includegraphics[width=0.45\textwidth]{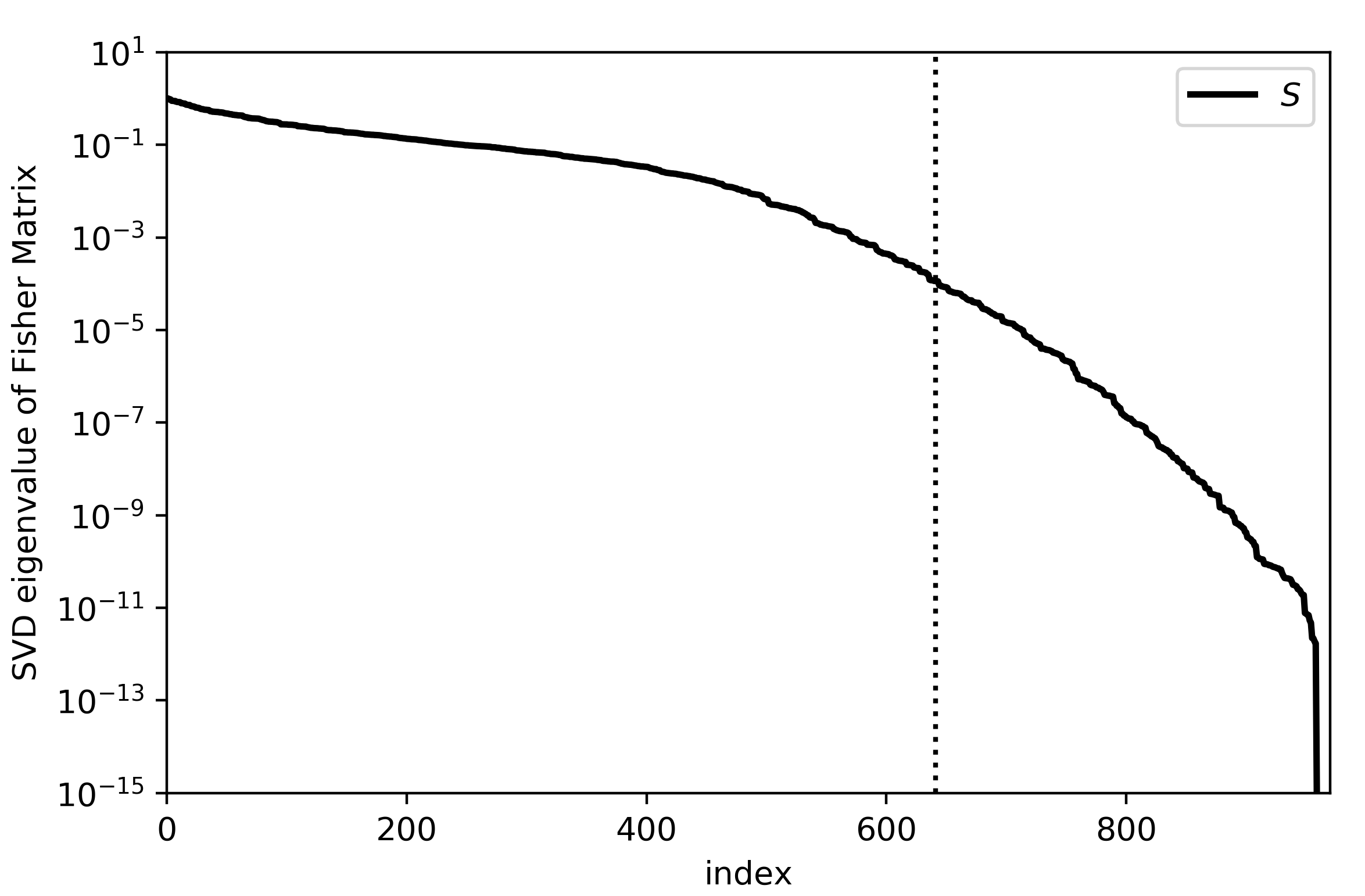}
    \caption{SVD eigenvalues of the Fisher information matrix $\Gamma_{lm,l'm'}$ are represented using the solid line. The dotted vertical line demarcates the eigenvalues considered for the inversion of regularized Fisher matrix $\Gamma'_{lm,l'm'}$. For demonstration, we have considered $l_{\rm{max}}=30$ (961 modes).}
    \label{pystoch_sph_fisher_eigenvalues}
\end{figure}

\begin{figure}
    \centering
    \includegraphics[width = 0.4\textwidth]{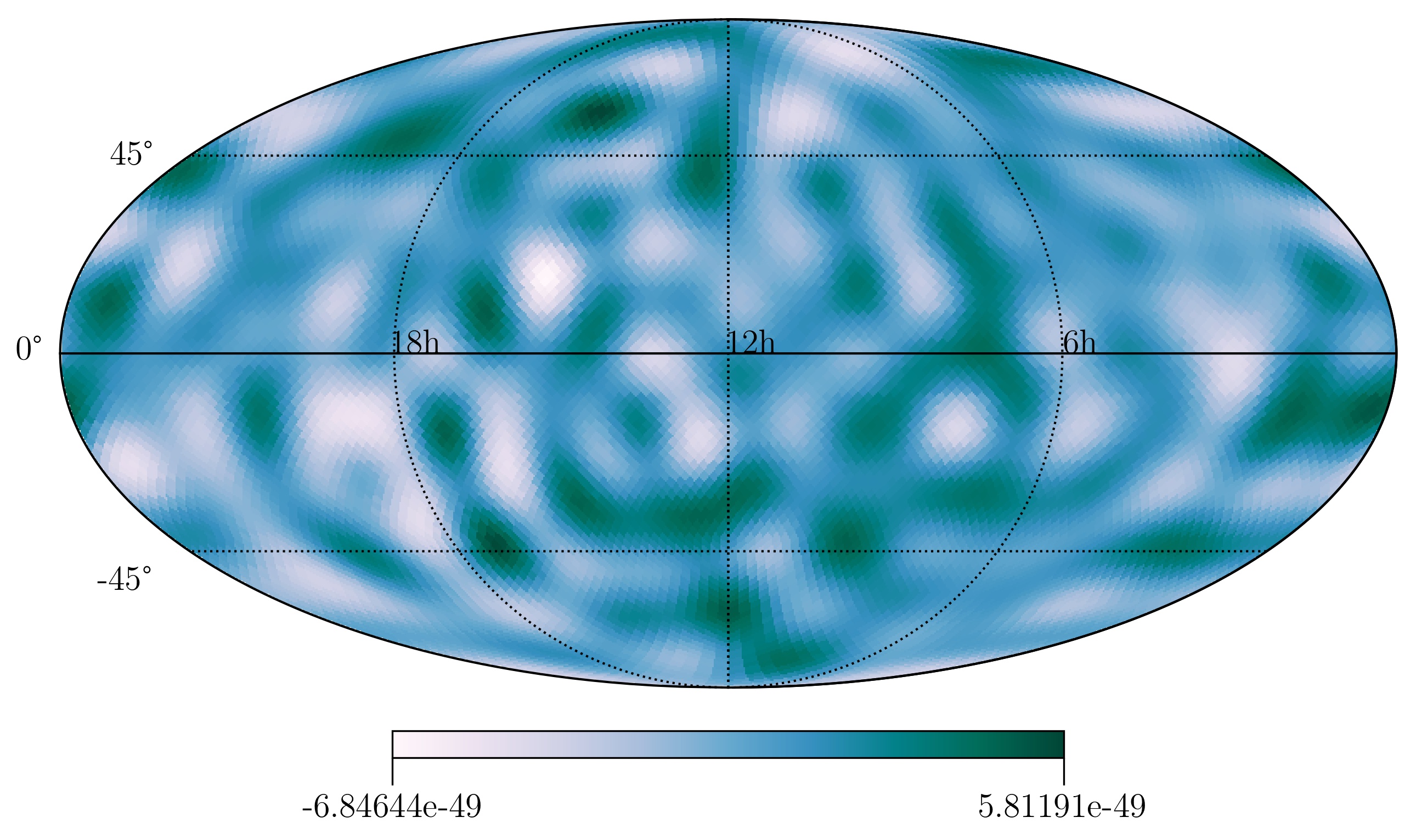}
    \caption{Clean map produced using the SVD regularized Fisher matrix is represented as a color bar plot on a Mollweide projection of the sky in ecliptic coordinates. These maps are obtained for a power-law spectral indices $\alpha = 3$ and contain information up to $l_{\rm{max}}=15$. }
    \label{pystoch_clean}
\end{figure}

Earlier studies utilized the SVD techniques to characterize~\cite{Mitra07} and to condition~\cite{Romano2017,Thrane09} the Fisher matrix. In this paper, as an example, we use the SVD technique to obtain the clean maps, even though  {\tt PyStoch} is capable of handling any type of regularized deconvolution~\cite{regDeconv}. The Fisher matrix, from its definition in Eq.~(\ref{eq_fisher}), is Hermitian, so its SVD takes the form
\begin{equation}
\label{fisher_svd}
\Gamma = U S U^*\,,
\end{equation}
where $U$ is a unitary matrix and $S$ is a diagonal matrix, whose nonzero elements are the positive and real eigenvalues of the Fisher matrix, arranged in descending order. To condition the matrix, a threshold $S_{\rm min}$ is chosen. The choice is made by considering the proper trade-off between the quality of the deconvolution and the increase in numerical noise from less sensitive modes. Any values below this cutoff are considered too small, and we replace them with infinity. Alternatively, they can be replaced with the smallest eigenvalue above the cut-off.

\begin{figure}[h]
    \centering
    \includegraphics[width = 0.45\textwidth]{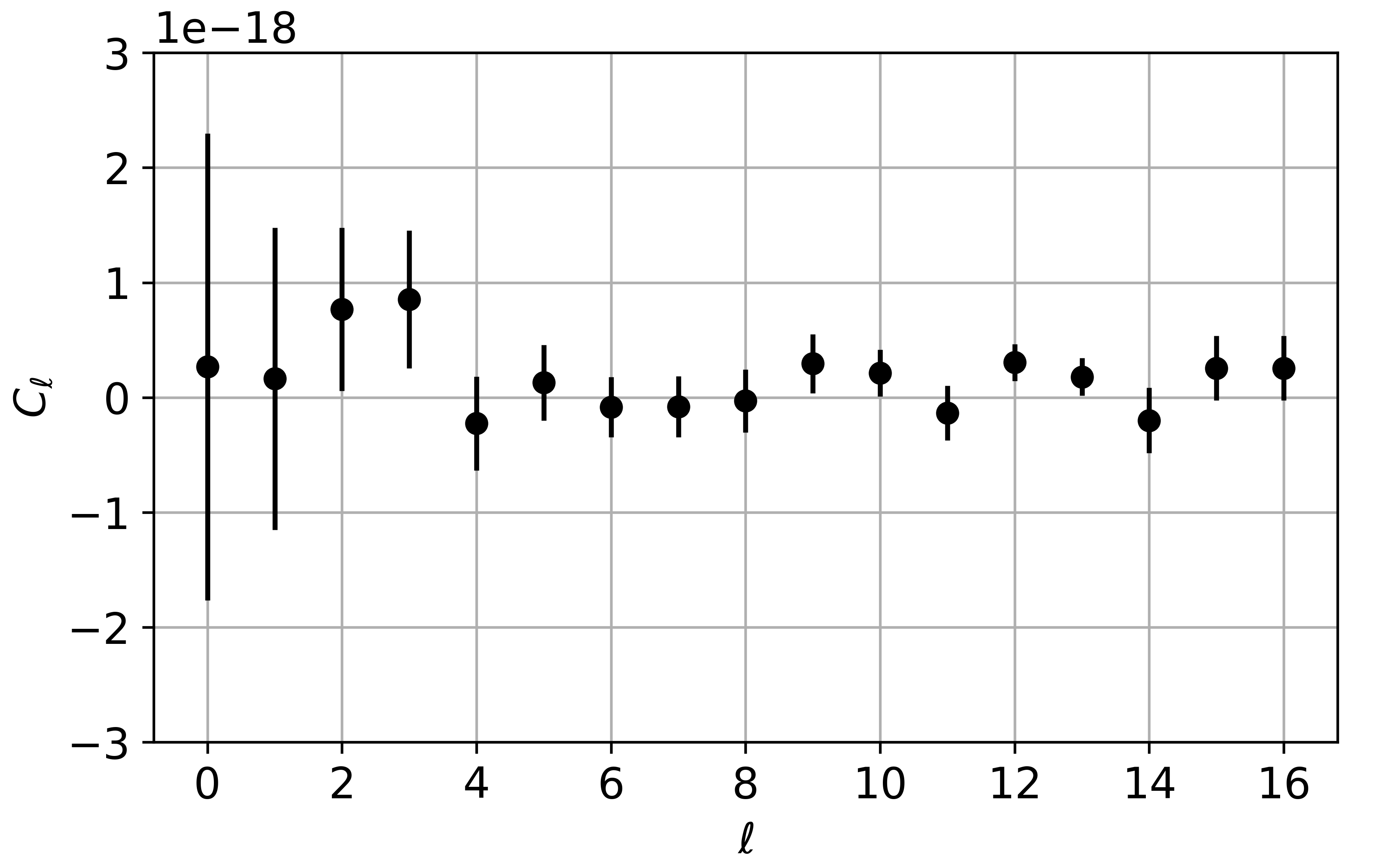}
    \caption{Estimate of the angular power spectrum, $C_l$, of the SGWB for a spectral shape characterized by $\alpha =3$.}
    \label{pystoch_Cl}
\end{figure}

In this demonstration, the singular value cutoff is chosen as $10^{-3}$ of the maximum eigenvalue. Any value below this threshold is padded by the $S_{\rm min}$. 

Now, one can easily write the inverse of regularized Fisher matrix, which is obtained using the modified $S'$ (see Fig.~\ref{pystoch_sph_fisher_eigenvalues}) as
\begin{equation}
\Gamma'^{-1}= U S'^{-1} U^*\,.
\end{equation}
By multiplying the inverted-regularized Fisher matrix with the dirty map, one can obtain the estimators of the spherical harmonic coefficients:
\begin{equation}
\mathcal{\hat{P}}_{lm} = \Gamma'^{-1}_{lm} X_{lm}.
\end{equation}
One can use the above clean map (Fig.\ref{pystoch_clean}) in the SpH basis, to construct the unbiased estimator of the angular power spectra, i.e.,
\begin{equation}
\label{ang_power_spectra}
\hat C_l = \frac{1}{2 l + 1} \, \sum_{m} \left[|\hat{\mathcal P}_{lm}|^2 - (\Gamma'^{-1})_{lm,lm}\right] \,.
\end{equation}
We have also computed the estimate of the angular power spectrum $C_l$ (describes angular scale of the structure found in the clean map), of the SGWB for a specific spectral distribution or signal model.  Figure~\ref{pystoch_Cl} shows the obtained $C_l$ characterized by spectral index $\alpha =3$ from the folded dataset. 
%
% - - - - -- - - - - - - - - - - - -  - - - - - - - - - - - - - - - - - - - 
%
%\bibliographystyle{apsrev4-2}
\bibliography{bib.bib}

\begin{thebibliography}{55}%
\makeatletter
\providecommand \@ifxundefined [1]{%
 \@ifx{#1\undefined}
}%
\providecommand \@ifnum [1]{%
 \ifnum #1\expandafter \@firstoftwo
 \else \expandafter \@secondoftwo
 \fi
}%
\providecommand \@ifx [1]{%
 \ifx #1\expandafter \@firstoftwo
 \else \expandafter \@secondoftwo
 \fi
}%
\providecommand \natexlab [1]{#1}%
\providecommand \enquote  [1]{``#1''}%
\providecommand \bibnamefont  [1]{#1}%
\providecommand \bibfnamefont [1]{#1}%
\providecommand \citenamefont [1]{#1}%
\providecommand \href@noop [0]{\@secondoftwo}%
\providecommand \href [0]{\begingroup \@sanitize@url \@href}%
\providecommand \@href[1]{\@@startlink{#1}\@@href}%
\providecommand \@@href[1]{\endgroup#1\@@endlink}%
\providecommand \@sanitize@url [0]{\catcode `\\12\catcode `\$12\catcode
  `\&12\catcode `\#12\catcode `\^12\catcode `\_12\catcode `\%12\relax}%
\providecommand \@@startlink[1]{}%
\providecommand \@@endlink[0]{}%
\providecommand \url  [0]{\begingroup\@sanitize@url \@url }%
\providecommand \@url [1]{\endgroup\@href {#1}{\urlprefix }}%
\providecommand \urlprefix  [0]{URL }%
\providecommand \Eprint [0]{\href }%
\providecommand \doibase [0]{http://dx.doi.org/}%
\providecommand \selectlanguage [0]{\@gobble}%
\providecommand \bibinfo  [0]{\@secondoftwo}%
\providecommand \bibfield  [0]{\@secondoftwo}%
\providecommand \translation [1]{[#1]}%
\providecommand \BibitemOpen [0]{}%
\providecommand \bibitemStop [0]{}%
\providecommand \bibitemNoStop [0]{.\EOS\space}%
\providecommand \EOS [0]{\spacefactor3000\relax}%
\providecommand \BibitemShut  [1]{\csname bibitem#1\endcsname}%
\let\auto@bib@innerbib\@empty
%</preamble>
\bibitem [{\citenamefont {Abbott}\ \emph {et~al.}(2016)\citenamefont {Abbott}
  \emph {et~al.}}]{GW150914}%
  \BibitemOpen
  \bibfield  {author} {\bibinfo {author} {\bibfnamefont {B.}~\bibnamefont
  {Abbott}} \emph {et~al.} (\bibinfo {collaboration} {LIGO Scientific,
  Virgo}),\ }\href {\doibase 10.1103/PhysRevLett.116.061102} {\bibfield
  {journal} {\bibinfo  {journal} {Phys. Rev. Lett.}\ }\textbf {\bibinfo
  {volume} {116}},\ \bibinfo {pages} {061102} (\bibinfo {year} {2016})},\
  \Eprint {http://arxiv.org/abs/1602.03837} {arXiv:1602.03837 [gr-qc]}
  \BibitemShut {NoStop}%
\bibitem [{\citenamefont {Abbott}\ \emph {et~al.}(2020)\citenamefont {Abbott}
  \emph {et~al.}}]{GWTC-2}%
  \BibitemOpen
  \bibfield  {author} {\bibinfo {author} {\bibfnamefont {R.}~\bibnamefont
  {Abbott}} \emph {et~al.} (\bibinfo {collaboration} {LIGO Scientific,
  Virgo}),\ }\href@noop {} {\  (\bibinfo {year} {2020})},\ \Eprint
  {http://arxiv.org/abs/2010.14527} {arXiv:2010.14527 [gr-qc]} \BibitemShut
  {NoStop}%
\bibitem [{\citenamefont {Punturo}\ \emph {et~al.}(2010)\citenamefont {Punturo}
  \emph {et~al.}}]{ET}%
  \BibitemOpen
  \bibfield  {author} {\bibinfo {author} {\bibfnamefont {M.}~\bibnamefont
  {Punturo}} \emph {et~al.},\ }\href {\doibase 10.1088/0264-9381/27/8/084007}
  {\bibfield  {journal} {\bibinfo  {journal} {Class. Quantum Gravity}\ }\textbf
  {\bibinfo {volume} {27}},\ \bibinfo {eid} {084007} (\bibinfo {year}
  {2010})}\BibitemShut {NoStop}%
\bibitem [{\citenamefont {Reitze}\ \emph {et~al.}(2019)\citenamefont {Reitze}
  \emph {et~al.}}]{CE}%
  \BibitemOpen
  \bibfield  {author} {\bibinfo {author} {\bibfnamefont {D.}~\bibnamefont
  {Reitze}} \emph {et~al.},\ }\href@noop {} {\bibfield  {journal} {\bibinfo
  {journal} {Bull. Am. Astron. Soc.}\ }\textbf {\bibinfo {volume} {51}},\
  \bibinfo {pages} {035} (\bibinfo {year} {2019})},\ \Eprint
  {http://arxiv.org/abs/1907.04833} {arXiv:1907.04833 [astro-ph.IM]}
  \BibitemShut {NoStop}%
\bibitem [{\citenamefont {{Amaro-Seoane}}\ \emph {et~al.}(2017)\citenamefont
  {{Amaro-Seoane}}, \citenamefont {{Audley}}, \citenamefont {{Babak}},
  \citenamefont {{Baker}}, \citenamefont {{Barausse}}, \citenamefont
  {{Bender}}, \citenamefont {{Berti}}, \citenamefont {{Binetruy}},
  \citenamefont {{Born}}, \citenamefont {{Bortoluzzi}}, \citenamefont {{Camp}},
  \citenamefont {{Caprini}}, \citenamefont {{Cardoso}}, \citenamefont
  {{Colpi}}, \citenamefont {{Conklin}}, \citenamefont {{Cornish}},
  \citenamefont {{Cutler}}, \citenamefont {{Danzmann}}, \citenamefont
  {{Dolesi}}, \citenamefont {{Ferraioli}}, \citenamefont {{Ferroni}},
  \citenamefont {{Fitzsimons}}, \citenamefont {{Gair}}, \citenamefont {{Gesa
  Bote}}, \citenamefont {{Giardini}}, \citenamefont {{Gibert}}, \citenamefont
  {{Grimani}}, \citenamefont {{Halloin}}, \citenamefont {{Heinzel}},
  \citenamefont {{Hertog}}, \citenamefont {{Hewitson}}, \citenamefont
  {{Holley-Bockelmann}}, \citenamefont {{Hollington}}, \citenamefont
  {{Hueller}}, \citenamefont {{Inchauspe}}, \citenamefont {{Jetzer}},
  \citenamefont {{Karnesis}}, \citenamefont {{Killow}}, \citenamefont
  {{Klein}}, \citenamefont {{Klipstein}}, \citenamefont {{Korsakova}},
  \citenamefont {{Larson}}, \citenamefont {{Livas}}, \citenamefont {{Lloro}},
  \citenamefont {{Man}}, \citenamefont {{Mance}}, \citenamefont {{Martino}},
  \citenamefont {{Mateos}}, \citenamefont {{McKenzie}}, \citenamefont
  {{McWilliams}}, \citenamefont {{Miller}}, \citenamefont {{Mueller}},
  \citenamefont {{Nardini}}, \citenamefont {{Nelemans}}, \citenamefont
  {{Nofrarias}}, \citenamefont {{Petiteau}}, \citenamefont {{Pivato}},
  \citenamefont {{Plagnol}}, \citenamefont {{Porter}}, \citenamefont
  {{Reiche}}, \citenamefont {{Robertson}}, \citenamefont {{Robertson}},
  \citenamefont {{Rossi}}, \citenamefont {{Russano}}, \citenamefont {{Schutz}},
  \citenamefont {{Sesana}}, \citenamefont {{Shoemaker}}, \citenamefont
  {{Slutsky}}, \citenamefont {{Sopuerta}}, \citenamefont {{Sumner}},
  \citenamefont {{Tamanini}}, \citenamefont {{Thorpe}}, \citenamefont
  {{Troebs}}, \citenamefont {{Vallisneri}}, \citenamefont {{Vecchio}},
  \citenamefont {{Vetrugno}}, \citenamefont {{Vitale}}, \citenamefont
  {{Volonteri}}, \citenamefont {{Wanner}}, \citenamefont {{Ward}},
  \citenamefont {{Wass}}, \citenamefont {{Weber}}, \citenamefont {{Ziemer}},\
  and\ \citenamefont {{Zweifel}}}]{LISA}%
  \BibitemOpen
  \bibfield  {author} {\bibinfo {author} {\bibfnamefont {P.}~\bibnamefont
  {{Amaro-Seoane}}}, \bibinfo {author} {\bibfnamefont {H.}~\bibnamefont
  {{Audley}}}, \bibinfo {author} {\bibfnamefont {S.}~\bibnamefont {{Babak}}},
  \bibinfo {author} {\bibfnamefont {J.}~\bibnamefont {{Baker}}}, \bibinfo
  {author} {\bibfnamefont {E.}~\bibnamefont {{Barausse}}}, \bibinfo {author}
  {\bibfnamefont {P.}~\bibnamefont {{Bender}}}, \bibinfo {author}
  {\bibfnamefont {E.}~\bibnamefont {{Berti}}}, \bibinfo {author} {\bibfnamefont
  {P.}~\bibnamefont {{Binetruy}}}, \bibinfo {author} {\bibfnamefont
  {M.}~\bibnamefont {{Born}}}, \bibinfo {author} {\bibfnamefont
  {D.}~\bibnamefont {{Bortoluzzi}}}, \bibinfo {author} {\bibfnamefont
  {J.}~\bibnamefont {{Camp}}}, \bibinfo {author} {\bibfnamefont
  {C.}~\bibnamefont {{Caprini}}}, \bibinfo {author} {\bibfnamefont
  {V.}~\bibnamefont {{Cardoso}}}, \bibinfo {author} {\bibfnamefont
  {M.}~\bibnamefont {{Colpi}}}, \bibinfo {author} {\bibfnamefont
  {J.}~\bibnamefont {{Conklin}}}, \bibinfo {author} {\bibfnamefont
  {N.}~\bibnamefont {{Cornish}}}, \bibinfo {author} {\bibfnamefont
  {C.}~\bibnamefont {{Cutler}}}, \bibinfo {author} {\bibfnamefont
  {K.}~\bibnamefont {{Danzmann}}}, \bibinfo {author} {\bibfnamefont
  {R.}~\bibnamefont {{Dolesi}}}, \bibinfo {author} {\bibfnamefont
  {L.}~\bibnamefont {{Ferraioli}}}, \bibinfo {author} {\bibfnamefont
  {V.}~\bibnamefont {{Ferroni}}}, \bibinfo {author} {\bibfnamefont
  {E.}~\bibnamefont {{Fitzsimons}}}, \bibinfo {author} {\bibfnamefont
  {J.}~\bibnamefont {{Gair}}}, \bibinfo {author} {\bibfnamefont
  {L.}~\bibnamefont {{Gesa Bote}}}, \bibinfo {author} {\bibfnamefont
  {D.}~\bibnamefont {{Giardini}}}, \bibinfo {author} {\bibfnamefont
  {F.}~\bibnamefont {{Gibert}}}, \bibinfo {author} {\bibfnamefont
  {C.}~\bibnamefont {{Grimani}}}, \bibinfo {author} {\bibfnamefont
  {H.}~\bibnamefont {{Halloin}}}, \bibinfo {author} {\bibfnamefont
  {G.}~\bibnamefont {{Heinzel}}}, \bibinfo {author} {\bibfnamefont
  {T.}~\bibnamefont {{Hertog}}}, \bibinfo {author} {\bibfnamefont
  {M.}~\bibnamefont {{Hewitson}}}, \bibinfo {author} {\bibfnamefont
  {K.}~\bibnamefont {{Holley-Bockelmann}}}, \bibinfo {author} {\bibfnamefont
  {D.}~\bibnamefont {{Hollington}}}, \bibinfo {author} {\bibfnamefont
  {M.}~\bibnamefont {{Hueller}}}, \bibinfo {author} {\bibfnamefont
  {H.}~\bibnamefont {{Inchauspe}}}, \bibinfo {author} {\bibfnamefont
  {P.}~\bibnamefont {{Jetzer}}}, \bibinfo {author} {\bibfnamefont
  {N.}~\bibnamefont {{Karnesis}}}, \bibinfo {author} {\bibfnamefont
  {C.}~\bibnamefont {{Killow}}}, \bibinfo {author} {\bibfnamefont
  {A.}~\bibnamefont {{Klein}}}, \bibinfo {author} {\bibfnamefont
  {B.}~\bibnamefont {{Klipstein}}}, \bibinfo {author} {\bibfnamefont
  {N.}~\bibnamefont {{Korsakova}}}, \bibinfo {author} {\bibfnamefont {S.~L.}\
  \bibnamefont {{Larson}}}, \bibinfo {author} {\bibfnamefont {J.}~\bibnamefont
  {{Livas}}}, \bibinfo {author} {\bibfnamefont {I.}~\bibnamefont {{Lloro}}},
  \bibinfo {author} {\bibfnamefont {N.}~\bibnamefont {{Man}}}, \bibinfo
  {author} {\bibfnamefont {D.}~\bibnamefont {{Mance}}}, \bibinfo {author}
  {\bibfnamefont {J.}~\bibnamefont {{Martino}}}, \bibinfo {author}
  {\bibfnamefont {I.}~\bibnamefont {{Mateos}}}, \bibinfo {author}
  {\bibfnamefont {K.}~\bibnamefont {{McKenzie}}}, \bibinfo {author}
  {\bibfnamefont {S.~T.}\ \bibnamefont {{McWilliams}}}, \bibinfo {author}
  {\bibfnamefont {C.}~\bibnamefont {{Miller}}}, \bibinfo {author}
  {\bibfnamefont {G.}~\bibnamefont {{Mueller}}}, \bibinfo {author}
  {\bibfnamefont {G.}~\bibnamefont {{Nardini}}}, \bibinfo {author}
  {\bibfnamefont {G.}~\bibnamefont {{Nelemans}}}, \bibinfo {author}
  {\bibfnamefont {M.}~\bibnamefont {{Nofrarias}}}, \bibinfo {author}
  {\bibfnamefont {A.}~\bibnamefont {{Petiteau}}}, \bibinfo {author}
  {\bibfnamefont {P.}~\bibnamefont {{Pivato}}}, \bibinfo {author}
  {\bibfnamefont {E.}~\bibnamefont {{Plagnol}}}, \bibinfo {author}
  {\bibfnamefont {E.}~\bibnamefont {{Porter}}}, \bibinfo {author}
  {\bibfnamefont {J.}~\bibnamefont {{Reiche}}}, \bibinfo {author}
  {\bibfnamefont {D.}~\bibnamefont {{Robertson}}}, \bibinfo {author}
  {\bibfnamefont {N.}~\bibnamefont {{Robertson}}}, \bibinfo {author}
  {\bibfnamefont {E.}~\bibnamefont {{Rossi}}}, \bibinfo {author} {\bibfnamefont
  {G.}~\bibnamefont {{Russano}}}, \bibinfo {author} {\bibfnamefont
  {B.}~\bibnamefont {{Schutz}}}, \bibinfo {author} {\bibfnamefont
  {A.}~\bibnamefont {{Sesana}}}, \bibinfo {author} {\bibfnamefont
  {D.}~\bibnamefont {{Shoemaker}}}, \bibinfo {author} {\bibfnamefont
  {J.}~\bibnamefont {{Slutsky}}}, \bibinfo {author} {\bibfnamefont {C.~F.}\
  \bibnamefont {{Sopuerta}}}, \bibinfo {author} {\bibfnamefont
  {T.}~\bibnamefont {{Sumner}}}, \bibinfo {author} {\bibfnamefont
  {N.}~\bibnamefont {{Tamanini}}}, \bibinfo {author} {\bibfnamefont
  {I.}~\bibnamefont {{Thorpe}}}, \bibinfo {author} {\bibfnamefont
  {M.}~\bibnamefont {{Troebs}}}, \bibinfo {author} {\bibfnamefont
  {M.}~\bibnamefont {{Vallisneri}}}, \bibinfo {author} {\bibfnamefont
  {A.}~\bibnamefont {{Vecchio}}}, \bibinfo {author} {\bibfnamefont
  {D.}~\bibnamefont {{Vetrugno}}}, \bibinfo {author} {\bibfnamefont
  {S.}~\bibnamefont {{Vitale}}}, \bibinfo {author} {\bibfnamefont
  {M.}~\bibnamefont {{Volonteri}}}, \bibinfo {author} {\bibfnamefont
  {G.}~\bibnamefont {{Wanner}}}, \bibinfo {author} {\bibfnamefont
  {H.}~\bibnamefont {{Ward}}}, \bibinfo {author} {\bibfnamefont
  {P.}~\bibnamefont {{Wass}}}, \bibinfo {author} {\bibfnamefont
  {W.}~\bibnamefont {{Weber}}}, \bibinfo {author} {\bibfnamefont
  {J.}~\bibnamefont {{Ziemer}}}, \ and\ \bibinfo {author} {\bibfnamefont
  {P.}~\bibnamefont {{Zweifel}}},\ }\href@noop {} {\bibfield  {journal}
  {\bibinfo  {journal} {arXiv e-prints}\ ,\ \bibinfo {eid} {arXiv:1702.00786}}
  (\bibinfo {year} {2017})},\ \Eprint {http://arxiv.org/abs/1702.00786}
  {arXiv:1702.00786 [astro-ph.IM]} \BibitemShut {NoStop}%
\bibitem [{\citenamefont {{Allen}}(1997)}]{1997rggr.conf..373A}%
  \BibitemOpen
  \bibfield  {author} {\bibinfo {author} {\bibfnamefont {B.}~\bibnamefont
  {{Allen}}},\ }in\ \href@noop {} {\emph {\bibinfo {booktitle} {Relativistic
  Gravitation and Gravitational Radiation}}},\ \bibinfo {editor} {edited by\
  \bibinfo {editor} {\bibfnamefont {J.-A.}\ \bibnamefont {{Marck}}}\ and\
  \bibinfo {editor} {\bibfnamefont {J.-P.}\ \bibnamefont {{Lasota}}}}\
  (\bibinfo {year} {1997})\ p.\ \bibinfo {pages} {373},\ \Eprint
  {http://arxiv.org/abs/gr-qc/9604033} {gr-qc/9604033} \BibitemShut {NoStop}%
\bibitem [{\citenamefont {Allen}\ and\ \citenamefont
  {Ottewill}(1997)}]{AllenOttewill}%
  \BibitemOpen
  \bibfield  {author} {\bibinfo {author} {\bibfnamefont {B.}~\bibnamefont
  {Allen}}\ and\ \bibinfo {author} {\bibfnamefont {A.~C.}\ \bibnamefont
  {Ottewill}},\ }\href {\doibase 10.1103/PhysRevD.56.545} {\bibfield  {journal}
  {\bibinfo  {journal} {Phys. Rev. D}\ }\textbf {\bibinfo {volume} {56}},\
  \bibinfo {pages} {545} (\bibinfo {year} {1997})}\BibitemShut {NoStop}%
\bibitem [{\citenamefont {{Abbott}}\ \emph {et~al.}(2018)\citenamefont
  {{Abbott}}, \citenamefont {{Abbott}}, \citenamefont {{Abbott}}, \citenamefont
  {{Acernese}}, \citenamefont {{Ackley}}, \citenamefont {{Adams}},
  \citenamefont {{Adams}}, \citenamefont {{Addesso}}, \citenamefont
  {{Adhikari}}, \citenamefont {{Adya}},\ and\ \citenamefont
  {et~al.}}]{2018PhRvL.120i1101A}%
  \BibitemOpen
  \bibfield  {author} {\bibinfo {author} {\bibfnamefont {B.~P.}\ \bibnamefont
  {{Abbott}}}, \bibinfo {author} {\bibfnamefont {R.}~\bibnamefont {{Abbott}}},
  \bibinfo {author} {\bibfnamefont {T.~D.}\ \bibnamefont {{Abbott}}}, \bibinfo
  {author} {\bibfnamefont {F.}~\bibnamefont {{Acernese}}}, \bibinfo {author}
  {\bibfnamefont {K.}~\bibnamefont {{Ackley}}}, \bibinfo {author}
  {\bibfnamefont {C.}~\bibnamefont {{Adams}}}, \bibinfo {author} {\bibfnamefont
  {T.}~\bibnamefont {{Adams}}}, \bibinfo {author} {\bibfnamefont
  {P.}~\bibnamefont {{Addesso}}}, \bibinfo {author} {\bibfnamefont {R.~X.}\
  \bibnamefont {{Adhikari}}}, \bibinfo {author} {\bibfnamefont {V.~B.}\
  \bibnamefont {{Adya}}}, \ and\ \bibinfo {author} {\bibnamefont {et~al.}},\
  }\href {\doibase 10.1103/PhysRevLett.120.091101} {\bibfield  {journal}
  {\bibinfo  {journal} {Physical Review Letters}\ }\textbf {\bibinfo {volume}
  {120}},\ \bibinfo {eid} {091101} (\bibinfo {year} {2018})},\ \Eprint
  {http://arxiv.org/abs/1710.05837} {arXiv:1710.05837 [gr-qc]} \BibitemShut
  {NoStop}%
\bibitem [{\citenamefont {{Rosado}}(2011)}]{2011PhRvD..84h4004R}%
  \BibitemOpen
  \bibfield  {author} {\bibinfo {author} {\bibfnamefont {P.~A.}\ \bibnamefont
  {{Rosado}}},\ }\href {\doibase 10.1103/PhysRevD.84.084004} {\bibfield
  {journal} {\bibinfo  {journal} {Phys. Rev. D}\ }\textbf {\bibinfo {volume}
  {84}},\ \bibinfo {eid} {084004} (\bibinfo {year} {2011})},\ \Eprint
  {http://arxiv.org/abs/1106.5795} {arXiv:1106.5795 [gr-qc]} \BibitemShut
  {NoStop}%
\bibitem [{\citenamefont {{Marassi}}\ \emph
  {et~al.}(2011{\natexlab{a}})\citenamefont {{Marassi}}, \citenamefont
  {{Schneider}}, \citenamefont {{Corvino}}, \citenamefont {{Ferrari}},\ and\
  \citenamefont {{Portegies Zwart}}}]{2011PhRvD..84l4037M}%
  \BibitemOpen
  \bibfield  {author} {\bibinfo {author} {\bibfnamefont {S.}~\bibnamefont
  {{Marassi}}}, \bibinfo {author} {\bibfnamefont {R.}~\bibnamefont
  {{Schneider}}}, \bibinfo {author} {\bibfnamefont {G.}~\bibnamefont
  {{Corvino}}}, \bibinfo {author} {\bibfnamefont {V.}~\bibnamefont
  {{Ferrari}}}, \ and\ \bibinfo {author} {\bibfnamefont {S.}~\bibnamefont
  {{Portegies Zwart}}},\ }\href {\doibase 10.1103/PhysRevD.84.124037}
  {\bibfield  {journal} {\bibinfo  {journal} {Phys. Rev. D}\ }\textbf {\bibinfo
  {volume} {84}},\ \bibinfo {eid} {124037} (\bibinfo {year}
  {2011}{\natexlab{a}})},\ \Eprint {http://arxiv.org/abs/1111.6125}
  {arXiv:1111.6125 [astro-ph.CO]} \BibitemShut {NoStop}%
\bibitem [{\citenamefont {{Zhu}}\ \emph {et~al.}(2011)\citenamefont {{Zhu}},
  \citenamefont {{Howell}}, \citenamefont {{Regimbau}}, \citenamefont
  {{Blair}},\ and\ \citenamefont {{Zhu}}}]{2011ApJ73986Z}%
  \BibitemOpen
  \bibfield  {author} {\bibinfo {author} {\bibfnamefont {X.-J.}\ \bibnamefont
  {{Zhu}}}, \bibinfo {author} {\bibfnamefont {E.}~\bibnamefont {{Howell}}},
  \bibinfo {author} {\bibfnamefont {T.}~\bibnamefont {{Regimbau}}}, \bibinfo
  {author} {\bibfnamefont {D.}~\bibnamefont {{Blair}}}, \ and\ \bibinfo
  {author} {\bibfnamefont {Z.-H.}\ \bibnamefont {{Zhu}}},\ }\href {\doibase
  10.1088/0004-637X/739/2/86} {\bibfield  {journal} {\bibinfo  {journal}
  {Astrophys. J.}\ }\textbf {\bibinfo {volume} {739}},\ \bibinfo {eid} {86}
  (\bibinfo {year} {2011})},\ \Eprint {http://arxiv.org/abs/1104.3565}
  {arXiv:1104.3565 [gr-qc]} \BibitemShut {NoStop}%
\bibitem [{\citenamefont {{Wu}}\ \emph {et~al.}(2012)\citenamefont {{Wu}},
  \citenamefont {{Mandic}},\ and\ \citenamefont
  {{Regimbau}}}]{2012PhRvD..85j4024W}%
  \BibitemOpen
  \bibfield  {author} {\bibinfo {author} {\bibfnamefont {C.}~\bibnamefont
  {{Wu}}}, \bibinfo {author} {\bibfnamefont {V.}~\bibnamefont {{Mandic}}}, \
  and\ \bibinfo {author} {\bibfnamefont {T.}~\bibnamefont {{Regimbau}}},\
  }\href {\doibase 10.1103/PhysRevD.85.104024} {\bibfield  {journal} {\bibinfo
  {journal} {Phys. Rev. D}\ }\textbf {\bibinfo {volume} {85}},\ \bibinfo {eid}
  {104024} (\bibinfo {year} {2012})},\ \Eprint {http://arxiv.org/abs/1112.1898}
  {arXiv:1112.1898 [gr-qc]} \BibitemShut {NoStop}%
\bibitem [{\citenamefont {{Zhu}}\ \emph {et~al.}(2013)\citenamefont {{Zhu}},
  \citenamefont {{Howell}}, \citenamefont {{Blair}},\ and\ \citenamefont
  {{Zhu}}}]{2013MNRAS.431..882Z}%
  \BibitemOpen
  \bibfield  {author} {\bibinfo {author} {\bibfnamefont {X.-J.}\ \bibnamefont
  {{Zhu}}}, \bibinfo {author} {\bibfnamefont {E.~J.}\ \bibnamefont {{Howell}}},
  \bibinfo {author} {\bibfnamefont {D.~G.}\ \bibnamefont {{Blair}}}, \ and\
  \bibinfo {author} {\bibfnamefont {Z.-H.}\ \bibnamefont {{Zhu}}},\ }\href
  {\doibase 10.1093/mnras/stt207} {\bibfield  {journal} {\bibinfo  {journal}
  {Mon. Not. Roy. Astron. Soc.}\ }\textbf {\bibinfo {volume} {431}},\ \bibinfo
  {pages} {882} (\bibinfo {year} {2013})},\ \Eprint
  {http://arxiv.org/abs/1209.0595} {arXiv:1209.0595 [gr-qc]} \BibitemShut
  {NoStop}%
\bibitem [{\citenamefont {{Dvorkin}}\ \emph {et~al.}(2016)\citenamefont
  {{Dvorkin}}, \citenamefont {{Uzan}}, \citenamefont {{Vangioni}},\ and\
  \citenamefont {{Silk}}}]{2016PhRvD..94j3011D}%
  \BibitemOpen
  \bibfield  {author} {\bibinfo {author} {\bibfnamefont {I.}~\bibnamefont
  {{Dvorkin}}}, \bibinfo {author} {\bibfnamefont {J.-P.}\ \bibnamefont
  {{Uzan}}}, \bibinfo {author} {\bibfnamefont {E.}~\bibnamefont {{Vangioni}}},
  \ and\ \bibinfo {author} {\bibfnamefont {J.}~\bibnamefont {{Silk}}},\ }\href
  {\doibase 10.1103/PhysRevD.94.103011} {\bibfield  {journal} {\bibinfo
  {journal} {Phys. Rev. D}\ }\textbf {\bibinfo {volume} {94}},\ \bibinfo {eid}
  {103011} (\bibinfo {year} {2016})},\ \Eprint
  {http://arxiv.org/abs/1607.06818} {arXiv:1607.06818 [astro-ph.HE]}
  \BibitemShut {NoStop}%
\bibitem [{\citenamefont {{P{\'e}rigois}}\ \emph {et~al.}(2021)\citenamefont
  {{P{\'e}rigois}}, \citenamefont {{Belczynski}}, \citenamefont {{Bulik}},\
  and\ \citenamefont {{Regimbau}}}]{2021PhRvD.103d3002P}%
  \BibitemOpen
  \bibfield  {author} {\bibinfo {author} {\bibfnamefont {C.}~\bibnamefont
  {{P{\'e}rigois}}}, \bibinfo {author} {\bibfnamefont {C.}~\bibnamefont
  {{Belczynski}}}, \bibinfo {author} {\bibfnamefont {T.}~\bibnamefont
  {{Bulik}}}, \ and\ \bibinfo {author} {\bibfnamefont {T.}~\bibnamefont
  {{Regimbau}}},\ }\href {\doibase 10.1103/PhysRevD.103.043002} {\bibfield
  {journal} {\bibinfo  {journal} {Phys. Rev. D}\ }\textbf {\bibinfo {volume}
  {103}},\ \bibinfo {eid} {043002} (\bibinfo {year} {2021})},\ \Eprint
  {http://arxiv.org/abs/2008.04890} {arXiv:2008.04890 [astro-ph.CO]}
  \BibitemShut {NoStop}%
\bibitem [{\citenamefont {{Dhurandhar}}\ \emph {et~al.}(2011)\citenamefont
  {{Dhurandhar}}, \citenamefont {{Tagoshi}}, \citenamefont {{Okada}},
  \citenamefont {{Kanda}},\ and\ \citenamefont {{Takahashi}}}]{hotspot}%
  \BibitemOpen
  \bibfield  {author} {\bibinfo {author} {\bibfnamefont {S.}~\bibnamefont
  {{Dhurandhar}}}, \bibinfo {author} {\bibfnamefont {H.}~\bibnamefont
  {{Tagoshi}}}, \bibinfo {author} {\bibfnamefont {Y.}~\bibnamefont {{Okada}}},
  \bibinfo {author} {\bibfnamefont {N.}~\bibnamefont {{Kanda}}}, \ and\
  \bibinfo {author} {\bibfnamefont {H.}~\bibnamefont {{Takahashi}}},\ }\href
  {\doibase 10.1103/PhysRevD.84.083007} {\bibfield  {journal} {\bibinfo
  {journal} {Phys. Rev. D}\ }\textbf {\bibinfo {volume} {84}},\ \bibinfo {eid}
  {083007} (\bibinfo {year} {2011})},\ \Eprint {http://arxiv.org/abs/1105.5842}
  {arXiv:1105.5842 [gr-qc]} \BibitemShut {NoStop}%
\bibitem [{\citenamefont {Hughes}(2014)}]{HUGHES201486}%
  \BibitemOpen
  \bibfield  {author} {\bibinfo {author} {\bibfnamefont {S.~A.}\ \bibnamefont
  {Hughes}},\ }\href {\doibase https://doi.org/10.1016/j.dark.2014.10.003}
  {\bibfield  {journal} {\bibinfo  {journal} {Physics of the Dark Universe}\
  }\textbf {\bibinfo {volume} {4}},\ \bibinfo {pages} {86 } (\bibinfo {year}
  {2014})},\ \bibinfo {note} {dARK TAUP2013}\BibitemShut {NoStop}%
\bibitem [{\citenamefont {Cusin}\ \emph {et~al.}(2019)\citenamefont {Cusin},
  \citenamefont {Durrer},\ and\ \citenamefont
  {Ferreira}}]{2018arXiv180710620C}%
  \BibitemOpen
  \bibfield  {author} {\bibinfo {author} {\bibfnamefont {G.}~\bibnamefont
  {Cusin}}, \bibinfo {author} {\bibfnamefont {R.}~\bibnamefont {Durrer}}, \
  and\ \bibinfo {author} {\bibfnamefont {P.~G.}\ \bibnamefont {Ferreira}},\
  }\href {\doibase 10.1103/PhysRevD.99.023534} {\bibfield  {journal} {\bibinfo
  {journal} {Phys. Rev. D}\ }\textbf {\bibinfo {volume} {99}},\ \bibinfo
  {pages} {023534} (\bibinfo {year} {2019})},\ \Eprint
  {http://arxiv.org/abs/1807.10620} {arXiv:1807.10620 [astro-ph.CO]}
  \BibitemShut {NoStop}%
\bibitem [{\citenamefont {Mingarelli}\ \emph {et~al.}(2013)\citenamefont
  {Mingarelli}, \citenamefont {Sidery}, \citenamefont {Mandel},\ and\
  \citenamefont {Vecchio}}]{PhysRevD.88.062005}%
  \BibitemOpen
  \bibfield  {author} {\bibinfo {author} {\bibfnamefont {C.~M.~F.}\
  \bibnamefont {Mingarelli}}, \bibinfo {author} {\bibfnamefont
  {T.}~\bibnamefont {Sidery}}, \bibinfo {author} {\bibfnamefont
  {I.}~\bibnamefont {Mandel}}, \ and\ \bibinfo {author} {\bibfnamefont
  {A.}~\bibnamefont {Vecchio}},\ }\href {\doibase 10.1103/PhysRevD.88.062005}
  {\bibfield  {journal} {\bibinfo  {journal} {Phys. Rev. D}\ }\textbf {\bibinfo
  {volume} {88}},\ \bibinfo {pages} {062005} (\bibinfo {year}
  {2013})}\BibitemShut {NoStop}%
\bibitem [{\citenamefont {Acernese}\ \emph {et~al.}(2015)\citenamefont
  {Acernese} \emph {et~al.}}]{virgo}%
  \BibitemOpen
  \bibfield  {author} {\bibinfo {author} {\bibfnamefont {F.}~\bibnamefont
  {Acernese}} \emph {et~al.} (\bibinfo {collaboration} {VIRGO}),\ }\href
  {\doibase 10.1088/0264-9381/32/2/024001} {\bibfield  {journal} {\bibinfo
  {journal} {Class. Quant. Grav.}\ }\textbf {\bibinfo {volume} {32}},\ \bibinfo
  {pages} {024001} (\bibinfo {year} {2015})},\ \Eprint
  {http://arxiv.org/abs/1408.3978} {arXiv:1408.3978 [gr-qc]} \BibitemShut
  {NoStop}%
\bibitem [{\citenamefont {Aso}\ \emph {et~al.}(2013)\citenamefont {Aso},
  \citenamefont {Michimura}, \citenamefont {Somiya}, \citenamefont {Ando},
  \citenamefont {Miyakawa}, \citenamefont {Sekiguchi}, \citenamefont
  {Tatsumi},\ and\ \citenamefont {Yamamoto}}]{Kagra}%
  \BibitemOpen
  \bibfield  {author} {\bibinfo {author} {\bibfnamefont {Y.}~\bibnamefont
  {Aso}}, \bibinfo {author} {\bibfnamefont {Y.}~\bibnamefont {Michimura}},
  \bibinfo {author} {\bibfnamefont {K.}~\bibnamefont {Somiya}}, \bibinfo
  {author} {\bibfnamefont {M.}~\bibnamefont {Ando}}, \bibinfo {author}
  {\bibfnamefont {O.}~\bibnamefont {Miyakawa}}, \bibinfo {author}
  {\bibfnamefont {T.}~\bibnamefont {Sekiguchi}}, \bibinfo {author}
  {\bibfnamefont {D.}~\bibnamefont {Tatsumi}}, \ and\ \bibinfo {author}
  {\bibfnamefont {H.}~\bibnamefont {Yamamoto}} (\bibinfo {collaboration}
  {KAGRA}),\ }\href {\doibase 10.1103/PhysRevD.88.043007} {\bibfield  {journal}
  {\bibinfo  {journal} {Phys. Rev. D}\ }\textbf {\bibinfo {volume} {88}},\
  \bibinfo {pages} {043007} (\bibinfo {year} {2013})},\ \Eprint
  {http://arxiv.org/abs/1306.6747} {arXiv:1306.6747 [gr-qc]} \BibitemShut
  {NoStop}%
\bibitem [{\citenamefont {Iyer}\ \emph {et~al.}(2011)\citenamefont {Iyer},
  \citenamefont {Souradeep}, \citenamefont {Unnikrishnan}, \citenamefont
  {Dhurandhar}, \citenamefont {Raja}, \citenamefont {Kumar},\ and\
  \citenamefont {Sengupta}}]{ligo_india}%
  \BibitemOpen
  \bibfield  {author} {\bibinfo {author} {\bibfnamefont {B.}~\bibnamefont
  {Iyer}}, \bibinfo {author} {\bibfnamefont {T.}~\bibnamefont {Souradeep}},
  \bibinfo {author} {\bibfnamefont {C.~S.}\ \bibnamefont {Unnikrishnan}},
  \bibinfo {author} {\bibfnamefont {S.}~\bibnamefont {Dhurandhar}}, \bibinfo
  {author} {\bibfnamefont {S.}~\bibnamefont {Raja}}, \bibinfo {author}
  {\bibfnamefont {A.}~\bibnamefont {Kumar}}, \ and\ \bibinfo {author}
  {\bibfnamefont {A.}~\bibnamefont {Sengupta}},\ }\href
  {https://dcc.ligo.org/M1100296/} {\bibfield  {journal} {\bibinfo  {journal}
  {LIGO-India Technical Report No. LIGO-M1100296}\ } (\bibinfo {year}
  {2011})}\BibitemShut {NoStop}%
\bibitem [{\citenamefont {Jenkins}\ and\ \citenamefont
  {Sakellariadou}(2018)}]{PhysRevD.98.063509}%
  \BibitemOpen
  \bibfield  {author} {\bibinfo {author} {\bibfnamefont {A.~C.}\ \bibnamefont
  {Jenkins}}\ and\ \bibinfo {author} {\bibfnamefont {M.}~\bibnamefont
  {Sakellariadou}},\ }\href {\doibase 10.1103/PhysRevD.98.063509} {\bibfield
  {journal} {\bibinfo  {journal} {Phys. Rev. D}\ }\textbf {\bibinfo {volume}
  {98}},\ \bibinfo {pages} {063509} (\bibinfo {year} {2018})}\BibitemShut
  {NoStop}%
\bibitem [{\citenamefont {{Mazumder}}\ \emph {et~al.}(2014)\citenamefont
  {{Mazumder}}, \citenamefont {{Mitra}},\ and\ \citenamefont
  {{Dhurandhar}}}]{2014PhRvD..89h4076M}%
  \BibitemOpen
  \bibfield  {author} {\bibinfo {author} {\bibfnamefont {N.}~\bibnamefont
  {{Mazumder}}}, \bibinfo {author} {\bibfnamefont {S.}~\bibnamefont {{Mitra}}},
  \ and\ \bibinfo {author} {\bibfnamefont {S.}~\bibnamefont {{Dhurandhar}}},\
  }\href {\doibase 10.1103/PhysRevD.89.084076} {\bibfield  {journal} {\bibinfo
  {journal} {Phys. Rev. D}\ }\textbf {\bibinfo {volume} {89}},\ \bibinfo {eid}
  {084076} (\bibinfo {year} {2014})},\ \Eprint {http://arxiv.org/abs/1401.5898}
  {arXiv:1401.5898 [gr-qc]} \BibitemShut {NoStop}%
\bibitem [{\citenamefont {{Jenkins}}\ \emph {et~al.}(2018)\citenamefont
  {{Jenkins}}, \citenamefont {{Sakellariadou}}, \citenamefont {{Regimbau}},\
  and\ \citenamefont {{Slezak}}}]{2018PhRvD..98f3501J}%
  \BibitemOpen
  \bibfield  {author} {\bibinfo {author} {\bibfnamefont {A.~C.}\ \bibnamefont
  {{Jenkins}}}, \bibinfo {author} {\bibfnamefont {M.}~\bibnamefont
  {{Sakellariadou}}}, \bibinfo {author} {\bibfnamefont {T.}~\bibnamefont
  {{Regimbau}}}, \ and\ \bibinfo {author} {\bibfnamefont {E.}~\bibnamefont
  {{Slezak}}},\ }\href {\doibase 10.1103/PhysRevD.98.063501} {\bibfield
  {journal} {\bibinfo  {journal} {Phys. Rev. D}\ }\textbf {\bibinfo {volume}
  {98}},\ \bibinfo {eid} {063501} (\bibinfo {year} {2018})},\ \Eprint
  {http://arxiv.org/abs/1806.01718} {arXiv:1806.01718 [astro-ph.CO]}
  \BibitemShut {NoStop}%
\bibitem [{\citenamefont {{Marassi}}\ \emph
  {et~al.}(2011{\natexlab{b}})\citenamefont {{Marassi}}, \citenamefont
  {{Ciolfi}}, \citenamefont {{Schneider}}, \citenamefont {{Stella}},\ and\
  \citenamefont {{Ferrari}}}]{2011MNRAS.411.2549M}%
  \BibitemOpen
  \bibfield  {author} {\bibinfo {author} {\bibfnamefont {S.}~\bibnamefont
  {{Marassi}}}, \bibinfo {author} {\bibfnamefont {R.}~\bibnamefont {{Ciolfi}}},
  \bibinfo {author} {\bibfnamefont {R.}~\bibnamefont {{Schneider}}}, \bibinfo
  {author} {\bibfnamefont {L.}~\bibnamefont {{Stella}}}, \ and\ \bibinfo
  {author} {\bibfnamefont {V.}~\bibnamefont {{Ferrari}}},\ }\href {\doibase
  10.1111/j.1365-2966.2010.17861.x} {\bibfield  {journal} {\bibinfo  {journal}
  {Mon. Not. Roy. Astron. Soc.}\ }\textbf {\bibinfo {volume} {411}},\ \bibinfo
  {pages} {2549} (\bibinfo {year} {2011}{\natexlab{b}})},\ \Eprint
  {http://arxiv.org/abs/1009.1240} {arXiv:1009.1240 [astro-ph.CO]} \BibitemShut
  {NoStop}%
\bibitem [{\citenamefont {{Rosado}}(2012)}]{2012PhRvD..86j4007R}%
  \BibitemOpen
  \bibfield  {author} {\bibinfo {author} {\bibfnamefont {P.~A.}\ \bibnamefont
  {{Rosado}}},\ }\href {\doibase 10.1103/PhysRevD.86.104007} {\bibfield
  {journal} {\bibinfo  {journal} {Phys. Rev. D}\ }\textbf {\bibinfo {volume}
  {86}},\ \bibinfo {eid} {104007} (\bibinfo {year} {2012})},\ \Eprint
  {http://arxiv.org/abs/1206.1330} {arXiv:1206.1330 [gr-qc]} \BibitemShut
  {NoStop}%
\bibitem [{\citenamefont {{Wu}}\ \emph {et~al.}(2013)\citenamefont {{Wu}},
  \citenamefont {{Mandic}},\ and\ \citenamefont
  {{Regimbau}}}]{2013PhRvD..87d2002W}%
  \BibitemOpen
  \bibfield  {author} {\bibinfo {author} {\bibfnamefont {C.-J.}\ \bibnamefont
  {{Wu}}}, \bibinfo {author} {\bibfnamefont {V.}~\bibnamefont {{Mandic}}}, \
  and\ \bibinfo {author} {\bibfnamefont {T.}~\bibnamefont {{Regimbau}}},\
  }\href {\doibase 10.1103/PhysRevD.87.042002} {\bibfield  {journal} {\bibinfo
  {journal} {Phys. Rev. D}\ }\textbf {\bibinfo {volume} {87}},\ \bibinfo {eid}
  {042002} (\bibinfo {year} {2013})}\BibitemShut {NoStop}%
\bibitem [{\citenamefont {{Lasky}}\ \emph {et~al.}(2013)\citenamefont
  {{Lasky}}, \citenamefont {{Bennett}},\ and\ \citenamefont
  {{Melatos}}}]{2013PhRvD..87f3004L}%
  \BibitemOpen
  \bibfield  {author} {\bibinfo {author} {\bibfnamefont {P.~D.}\ \bibnamefont
  {{Lasky}}}, \bibinfo {author} {\bibfnamefont {M.~F.}\ \bibnamefont
  {{Bennett}}}, \ and\ \bibinfo {author} {\bibfnamefont {A.}~\bibnamefont
  {{Melatos}}},\ }\href {\doibase 10.1103/PhysRevD.87.063004} {\bibfield
  {journal} {\bibinfo  {journal} {Phys. Rev. D}\ }\textbf {\bibinfo {volume}
  {87}},\ \bibinfo {eid} {063004} (\bibinfo {year} {2013})},\ \Eprint
  {http://arxiv.org/abs/1302.6033} {arXiv:1302.6033 [astro-ph.HE]} \BibitemShut
  {NoStop}%
\bibitem [{\citenamefont {Arzoumanian}\ \emph {et~al.}(2018)\citenamefont
  {Arzoumanian} \emph {et~al.}}]{Arzoumanian_2018}%
  \BibitemOpen
  \bibfield  {author} {\bibinfo {author} {\bibfnamefont {Z.}~\bibnamefont
  {Arzoumanian}} \emph {et~al.} (\bibinfo {collaboration} {NANOGRAV}),\ }\href
  {\doibase 10.3847/1538-4357/aabd3b} {\bibfield  {journal} {\bibinfo
  {journal} {Astrophys. J.}\ }\textbf {\bibinfo {volume} {859}},\ \bibinfo
  {pages} {47} (\bibinfo {year} {2018})},\ \Eprint
  {http://arxiv.org/abs/1801.02617} {arXiv:1801.02617 [astro-ph.HE]}
  \BibitemShut {NoStop}%
\bibitem [{\citenamefont {Bian}\ \emph {et~al.}(2020)\citenamefont {Bian},
  \citenamefont {Liu},\ and\ \citenamefont {Zhou}}]{NANOgrav}%
  \BibitemOpen
  \bibfield  {author} {\bibinfo {author} {\bibfnamefont {L.}~\bibnamefont
  {Bian}}, \bibinfo {author} {\bibfnamefont {J.}~\bibnamefont {Liu}}, \ and\
  \bibinfo {author} {\bibfnamefont {R.}~\bibnamefont {Zhou}},\ }\href@noop {}
  {\  (\bibinfo {year} {2020})},\ \Eprint {http://arxiv.org/abs/2009.13893}
  {arXiv:2009.13893 [astro-ph.CO]} \BibitemShut {NoStop}%
\bibitem [{\citenamefont {Romano}\ and\ \citenamefont
  {Cornish}(2017)}]{Romano2017}%
  \BibitemOpen
  \bibfield  {author} {\bibinfo {author} {\bibfnamefont {J.~D.}\ \bibnamefont
  {Romano}}\ and\ \bibinfo {author} {\bibfnamefont {N.~J.}\ \bibnamefont
  {Cornish}},\ }\href {\doibase 10.1007/s41114-017-0004-1} {\bibfield
  {journal} {\bibinfo  {journal} {Living Rev. Rel.}\ }\textbf {\bibinfo
  {volume} {20}},\ \bibinfo {pages} {2} (\bibinfo {year} {2017})},\ \Eprint
  {http://arxiv.org/abs/1608.06889} {arXiv:1608.06889 [gr-qc]} \BibitemShut
  {NoStop}%
\bibitem [{\citenamefont {Michelson}(1987)}]{Michelson87}%
  \BibitemOpen
  \bibfield  {author} {\bibinfo {author} {\bibfnamefont {P.~F.}\ \bibnamefont
  {Michelson}},\ }\href@noop {} {\bibfield  {journal} {\bibinfo  {journal}
  {Monthly Notices of the Royal Astronomical Society}\ }\textbf {\bibinfo
  {volume} {227}},\ \bibinfo {pages} {933} (\bibinfo {year}
  {1987})}\BibitemShut {NoStop}%
\bibitem [{\citenamefont {Christensen}(1992)}]{christ92}%
  \BibitemOpen
  \bibfield  {author} {\bibinfo {author} {\bibfnamefont {N.}~\bibnamefont
  {Christensen}},\ }\href {\doibase 10.1103/PhysRevD.46.5250} {\bibfield
  {journal} {\bibinfo  {journal} {Phys. Rev. D}\ }\textbf {\bibinfo {volume}
  {46}},\ \bibinfo {pages} {5250} (\bibinfo {year} {1992})}\BibitemShut
  {NoStop}%
\bibitem [{\citenamefont {Flanagan}(1993)}]{flan93}%
  \BibitemOpen
  \bibfield  {author} {\bibinfo {author} {\bibfnamefont {E.~E.}\ \bibnamefont
  {Flanagan}},\ }\href {\doibase 10.1103/PhysRevD.48.2389} {\bibfield
  {journal} {\bibinfo  {journal} {Phys. Rev. D}\ }\textbf {\bibinfo {volume}
  {48}},\ \bibinfo {pages} {2389} (\bibinfo {year} {1993})}\BibitemShut
  {NoStop}%
\bibitem [{\citenamefont {Allen}\ and\ \citenamefont {Romano}(1999)}]{allen01}%
  \BibitemOpen
  \bibfield  {author} {\bibinfo {author} {\bibfnamefont {B.}~\bibnamefont
  {Allen}}\ and\ \bibinfo {author} {\bibfnamefont {J.~D.}\ \bibnamefont
  {Romano}},\ }\href@noop {} {\bibfield  {journal} {\bibinfo  {journal}
  {Physical Review}\ }\textbf {\bibinfo {volume} {D59}},\ \bibinfo {pages}
  {102001} (\bibinfo {year} {1999})},\ \Eprint
  {http://arxiv.org/abs/gr-qc/9710117} {gr-qc/9710117} \BibitemShut {NoStop}%
\bibitem [{\citenamefont {Lazzarini}\ and\ \citenamefont
  {Romano}(2004)}]{LazzariniRomano}%
  \BibitemOpen
  \bibfield  {author} {\bibinfo {author} {\bibfnamefont {A.}~\bibnamefont
  {Lazzarini}}\ and\ \bibinfo {author} {\bibfnamefont {J.}~\bibnamefont
  {Romano}},\ }\href@noop {} {\emph {\bibinfo {title} {Use of overlapping
  windows in the stochastic background search}}},\ \bibinfo {type} {Internal
  working note}\ \bibinfo {number} {LIGO-T040089-00-Z}\ (\bibinfo
  {institution} {{Laser Interferometer Gravitational Wave Observatory
  (LIGO)}},\ \bibinfo {year} {2004})\BibitemShut {NoStop}%
\bibitem [{\citenamefont {Alonso}\ \emph {et~al.}(2020)\citenamefont {Alonso},
  \citenamefont {Contaldi}, \citenamefont {Cusin}, \citenamefont {Ferreira},\
  and\ \citenamefont {Renzini}}]{Alonso_Contaldi}%
  \BibitemOpen
  \bibfield  {author} {\bibinfo {author} {\bibfnamefont {D.}~\bibnamefont
  {Alonso}}, \bibinfo {author} {\bibfnamefont {C.~R.}\ \bibnamefont
  {Contaldi}}, \bibinfo {author} {\bibfnamefont {G.}~\bibnamefont {Cusin}},
  \bibinfo {author} {\bibfnamefont {P.~G.}\ \bibnamefont {Ferreira}}, \ and\
  \bibinfo {author} {\bibfnamefont {A.~I.}\ \bibnamefont {Renzini}},\ }\href
  {\doibase 10.1103/PhysRevD.101.124048} {\bibfield  {journal} {\bibinfo
  {journal} {Phys. Rev. D}\ }\textbf {\bibinfo {volume} {101}},\ \bibinfo
  {pages} {124048} (\bibinfo {year} {2020})}\BibitemShut {NoStop}%
\bibitem [{\citenamefont {Ballmer}(2006)}]{ballmer06}%
  \BibitemOpen
  \bibfield  {author} {\bibinfo {author} {\bibfnamefont {S.~W.}\ \bibnamefont
  {Ballmer}},\ }\href {\doibase 10.1088/0264-9381/23/8/S23} {\bibfield
  {journal} {\bibinfo  {journal} {Class. Quant. Grav.}\ }\textbf {\bibinfo
  {volume} {23}},\ \bibinfo {pages} {S179} (\bibinfo {year} {2006})},\ \Eprint
  {http://arxiv.org/abs/gr-qc/0510096} {arXiv:gr-qc/0510096 [gr-qc]}
  \BibitemShut {NoStop}%
%%CITATION = GR-QC/0510096;%%
\bibitem [{\citenamefont {Mitra}\ \emph {et~al.}(2008)\citenamefont {Mitra},
  \citenamefont {Dhurandhar}, \citenamefont {Souradeep}, \citenamefont
  {Lazzarini}, \citenamefont {Mandic} \emph {et~al.}}]{Mitra07}%
  \BibitemOpen
  \bibfield  {author} {\bibinfo {author} {\bibfnamefont {S.}~\bibnamefont
  {Mitra}}, \bibinfo {author} {\bibfnamefont {S.}~\bibnamefont {Dhurandhar}},
  \bibinfo {author} {\bibfnamefont {T.}~\bibnamefont {Souradeep}}, \bibinfo
  {author} {\bibfnamefont {A.}~\bibnamefont {Lazzarini}}, \bibinfo {author}
  {\bibfnamefont {V.}~\bibnamefont {Mandic}},  \emph {et~al.},\ }\href
  {\doibase 10.1103/PhysRevD.77.042002} {\bibfield  {journal} {\bibinfo
  {journal} {Phys.Rev.}\ }\textbf {\bibinfo {volume} {D77}},\ \bibinfo {pages}
  {042002} (\bibinfo {year} {2008})},\ \Eprint {http://arxiv.org/abs/0708.2728}
  {arXiv:0708.2728 [gr-qc]} \BibitemShut {NoStop}%
%%CITATION = ARXIV:0708.2728;%%
\bibitem [{\citenamefont {Thrane}\ \emph {et~al.}(2009)\citenamefont {Thrane},
  \citenamefont {Ballmer}, \citenamefont {Romano}, \citenamefont {Mitra},
  \citenamefont {Talukder}, \citenamefont {Bose},\ and\ \citenamefont
  {Mandic}}]{Thrane09}%
  \BibitemOpen
  \bibfield  {author} {\bibinfo {author} {\bibfnamefont {E.}~\bibnamefont
  {Thrane}}, \bibinfo {author} {\bibfnamefont {S.}~\bibnamefont {Ballmer}},
  \bibinfo {author} {\bibfnamefont {J.~D.}\ \bibnamefont {Romano}}, \bibinfo
  {author} {\bibfnamefont {S.}~\bibnamefont {Mitra}}, \bibinfo {author}
  {\bibfnamefont {D.}~\bibnamefont {Talukder}}, \bibinfo {author}
  {\bibfnamefont {S.}~\bibnamefont {Bose}}, \ and\ \bibinfo {author}
  {\bibfnamefont {V.}~\bibnamefont {Mandic}},\ }\href {\doibase
  10.1103/PhysRevD.80.122002} {\bibfield  {journal} {\bibinfo  {journal} {Phys.
  Rev.}\ }\textbf {\bibinfo {volume} {D80}},\ \bibinfo {pages} {122002}
  (\bibinfo {year} {2009})},\ \Eprint {http://arxiv.org/abs/0910.0858}
  {arXiv:0910.0858 [astro-ph.IM]} \BibitemShut {NoStop}%
%%CITATION = ARXIV:0910.0858;%%
\bibitem [{\citenamefont {Calore}\ \emph {et~al.}(2019)\citenamefont {Calore},
  \citenamefont {Regimbau},\ and\ \citenamefont {Serpico}}]{fermiLAT}%
  \BibitemOpen
  \bibfield  {author} {\bibinfo {author} {\bibfnamefont {F.}~\bibnamefont
  {Calore}}, \bibinfo {author} {\bibfnamefont {T.}~\bibnamefont {Regimbau}}, \
  and\ \bibinfo {author} {\bibfnamefont {P.~D.}\ \bibnamefont {Serpico}},\
  }\href {\doibase 10.1103/PhysRevLett.122.081103} {\bibfield  {journal}
  {\bibinfo  {journal} {Phys. Rev. Lett.}\ }\textbf {\bibinfo {volume} {122}},\
  \bibinfo {pages} {081103} (\bibinfo {year} {2019})},\ \Eprint
  {http://arxiv.org/abs/1812.05094} {arXiv:1812.05094 [astro-ph.HE]}
  \BibitemShut {NoStop}%
\bibitem [{\citenamefont {Cusin}\ \emph {et~al.}(2017)\citenamefont {Cusin},
  \citenamefont {Pitrou},\ and\ \citenamefont {Uzan}}]{Cusin:2017fwz}%
  \BibitemOpen
  \bibfield  {author} {\bibinfo {author} {\bibfnamefont {G.}~\bibnamefont
  {Cusin}}, \bibinfo {author} {\bibfnamefont {C.}~\bibnamefont {Pitrou}}, \
  and\ \bibinfo {author} {\bibfnamefont {J.-P.}\ \bibnamefont {Uzan}},\ }\href
  {\doibase 10.1103/PhysRevD.96.103019} {\bibfield  {journal} {\bibinfo
  {journal} {Phys. Rev. D}\ }\textbf {\bibinfo {volume} {96}},\ \bibinfo
  {pages} {103019} (\bibinfo {year} {2017})},\ \Eprint
  {http://arxiv.org/abs/1704.06184} {arXiv:1704.06184 [astro-ph.CO]}
  \BibitemShut {NoStop}%
\bibitem [{\citenamefont {Talukder}\ \emph {et~al.}(2011)\citenamefont
  {Talukder}, \citenamefont {Mitra},\ and\ \citenamefont
  {Bose}}]{Talukder:2010yd}%
  \BibitemOpen
  \bibfield  {author} {\bibinfo {author} {\bibfnamefont {D.}~\bibnamefont
  {Talukder}}, \bibinfo {author} {\bibfnamefont {S.}~\bibnamefont {Mitra}}, \
  and\ \bibinfo {author} {\bibfnamefont {S.}~\bibnamefont {Bose}},\ }\href
  {\doibase 10.1103/PhysRevD.83.063002} {\bibfield  {journal} {\bibinfo
  {journal} {Phys. Rev. D}\ }\textbf {\bibinfo {volume} {83}},\ \bibinfo
  {pages} {063002} (\bibinfo {year} {2011})},\ \Eprint
  {http://arxiv.org/abs/1012.4530} {arXiv:1012.4530 [gr-qc]} \BibitemShut
  {NoStop}%
\bibitem [{\citenamefont {Renzini}\ and\ \citenamefont
  {Contaldi}(2019)}]{Renzini}%
  \BibitemOpen
  \bibfield  {author} {\bibinfo {author} {\bibfnamefont {A.~I.}\ \bibnamefont
  {Renzini}}\ and\ \bibinfo {author} {\bibfnamefont {C.~R.}\ \bibnamefont
  {Contaldi}},\ }\href {\doibase 10.1103/PhysRevD.100.063527} {\bibfield
  {journal} {\bibinfo  {journal} {Phys. Rev. D}\ }\textbf {\bibinfo {volume}
  {100}},\ \bibinfo {pages} {063527} (\bibinfo {year} {2019})}\BibitemShut
  {NoStop}%
\bibitem [{\citenamefont {Ain}\ \emph {et~al.}(2015)\citenamefont {Ain},
  \citenamefont {Dalvi},\ and\ \citenamefont {Mitra}}]{Ain_Folding}%
  \BibitemOpen
  \bibfield  {author} {\bibinfo {author} {\bibfnamefont {A.}~\bibnamefont
  {Ain}}, \bibinfo {author} {\bibfnamefont {P.}~\bibnamefont {Dalvi}}, \ and\
  \bibinfo {author} {\bibfnamefont {S.}~\bibnamefont {Mitra}},\ }\href
  {\doibase 10.1103/PhysRevD.92.022003} {\bibfield  {journal} {\bibinfo
  {journal} {Phys. Rev.}\ }\textbf {\bibinfo {volume} {D92}},\ \bibinfo {pages}
  {022003} (\bibinfo {year} {2015})},\ \Eprint
  {http://arxiv.org/abs/1504.01714} {arXiv:1504.01714 [gr-qc]} \BibitemShut
  {NoStop}%
%%CITATION = ARXIV:1504.01714;%%
\bibitem [{\citenamefont {Ain}\ \emph {et~al.}(2018)\citenamefont {Ain},
  \citenamefont {Suresh},\ and\ \citenamefont {Mitra}}]{pystoch}%
  \BibitemOpen
  \bibfield  {author} {\bibinfo {author} {\bibfnamefont {A.}~\bibnamefont
  {Ain}}, \bibinfo {author} {\bibfnamefont {J.}~\bibnamefont {Suresh}}, \ and\
  \bibinfo {author} {\bibfnamefont {S.}~\bibnamefont {Mitra}},\ }\href
  {\doibase 10.1103/PhysRevD.98.024001} {\bibfield  {journal} {\bibinfo
  {journal} {Phys. Rev.}\ }\textbf {\bibinfo {volume} {D98}},\ \bibinfo {pages}
  {024001} (\bibinfo {year} {2018})},\ \Eprint
  {http://arxiv.org/abs/1803.08285} {arXiv:1803.08285 [gr-qc]} \BibitemShut
  {NoStop}%
%%CITATION = ARXIV:1803.08285;%%
\bibitem [{\citenamefont {Abbott}\ \emph {et~al.}(2021)\citenamefont {Abbott}
  \emph {et~al.}}]{O3-BBR}%
  \BibitemOpen
  \bibfield  {author} {\bibinfo {author} {\bibfnamefont {R.}~\bibnamefont
  {Abbott}} \emph {et~al.} (\bibinfo {collaboration} {LIGO Scientific, Virgo,
  KAGRA}),\ }\href@noop {} {\  (\bibinfo {year} {2021})},\ \Eprint
  {http://arxiv.org/abs/2103.08520} {arXiv:2103.08520 [gr-qc]} \BibitemShut
  {NoStop}%
\bibitem [{\citenamefont {Gorski}\ \emph {et~al.}(2005)\citenamefont {Gorski},
  \citenamefont {Hivon}, \citenamefont {Banday}, \citenamefont {Wandelt},
  \citenamefont {Hansen}, \citenamefont {Reinecke},\ and\ \citenamefont
  {Bartelman}}]{HEALPix}%
  \BibitemOpen
  \bibfield  {author} {\bibinfo {author} {\bibfnamefont {K.~M.}\ \bibnamefont
  {Gorski}}, \bibinfo {author} {\bibfnamefont {E.}~\bibnamefont {Hivon}},
  \bibinfo {author} {\bibfnamefont {A.~J.}\ \bibnamefont {Banday}}, \bibinfo
  {author} {\bibfnamefont {B.~D.}\ \bibnamefont {Wandelt}}, \bibinfo {author}
  {\bibfnamefont {F.~K.}\ \bibnamefont {Hansen}}, \bibinfo {author}
  {\bibfnamefont {M.}~\bibnamefont {Reinecke}}, \ and\ \bibinfo {author}
  {\bibfnamefont {M.}~\bibnamefont {Bartelman}},\ }\href {\doibase
  10.1086/427976} {\bibfield  {journal} {\bibinfo  {journal} {Astrophys. J.}\
  }\textbf {\bibinfo {volume} {622}},\ \bibinfo {pages} {759} (\bibinfo {year}
  {2005})},\ \Eprint {http://arxiv.org/abs/astro-ph/0409513}
  {arXiv:astro-ph/0409513 [astro-ph]} \BibitemShut {NoStop}%
%%CITATION = ASTRO-PH/0409513;%%
\bibitem [{\citenamefont {Abbott}\ \emph {et~al.}(2017)\citenamefont {Abbott}
  \emph {et~al.}}]{O1directional}%
  \BibitemOpen
  \bibfield  {author} {\bibinfo {author} {\bibfnamefont {B.~P.}\ \bibnamefont
  {Abbott}} \emph {et~al.} (\bibinfo {collaboration} {LIGO Scientific,
  Virgo}),\ }\href {\doibase 10.1103/PhysRevLett.118.121102} {\bibfield
  {journal} {\bibinfo  {journal} {Phys. Rev. Lett.}\ }\textbf {\bibinfo
  {volume} {118}},\ \bibinfo {pages} {121102} (\bibinfo {year} {2017})},\
  \Eprint {http://arxiv.org/abs/1612.02030} {arXiv:1612.02030 [gr-qc]}
  \BibitemShut {NoStop}%
\bibitem [{\citenamefont {Abbott}\ \emph {et~al.}(2019)\citenamefont {Abbott}
  \emph {et~al.}}]{O2directional}%
  \BibitemOpen
  \bibfield  {author} {\bibinfo {author} {\bibfnamefont {B.}~\bibnamefont
  {Abbott}} \emph {et~al.} (\bibinfo {collaboration} {LIGO Scientific,
  Virgo}),\ }\href {\doibase 10.1103/PhysRevD.100.062001} {\bibfield  {journal}
  {\bibinfo  {journal} {Phys. Rev. D}\ }\textbf {\bibinfo {volume} {100}},\
  \bibinfo {pages} {062001} (\bibinfo {year} {2019})},\ \Eprint
  {http://arxiv.org/abs/1903.08844} {arXiv:1903.08844 [gr-qc]} \BibitemShut
  {NoStop}%
\bibitem [{\citenamefont {Jackson}(1998)}]{jackson}%
  \BibitemOpen
  \bibfield  {author} {\bibinfo {author} {\bibfnamefont {J.~D.}\ \bibnamefont
  {Jackson}},\ }\href@noop {} {\emph {\bibinfo {title} {{Classical
  Electrodynamics}}}}\ (\bibinfo  {publisher} {Wiley},\ \bibinfo {year}
  {1998})\BibitemShut {NoStop}%
\bibitem [{\citenamefont {Finn}\ \emph {et~al.}(2009)\citenamefont {Finn},
  \citenamefont {Larson},\ and\ \citenamefont {Romano}}]{ORF_Finn}%
  \BibitemOpen
  \bibfield  {author} {\bibinfo {author} {\bibfnamefont {L.~S.}\ \bibnamefont
  {Finn}}, \bibinfo {author} {\bibfnamefont {S.~L.}\ \bibnamefont {Larson}}, \
  and\ \bibinfo {author} {\bibfnamefont {J.~D.}\ \bibnamefont {Romano}},\
  }\href {\doibase 10.1103/PhysRevD.79.062003} {\bibfield  {journal} {\bibinfo
  {journal} {Phys. Rev.}\ }\textbf {\bibinfo {volume} {D79}},\ \bibinfo {pages}
  {062003} (\bibinfo {year} {2009})},\ \Eprint {http://arxiv.org/abs/0811.3582}
  {arXiv:0811.3582 [gr-qc]} \BibitemShut {NoStop}%
%%CITATION = ARXIV:0811.3582;%%
\bibitem [{git()}]{gitpublic}%
  \BibitemOpen
  \href@noop {} {}\bibinfo {howpublished}
  {\url{https://git.ligo.org/stochastic-public/stochastic}}\BibitemShut
  {NoStop}%
\bibitem [{\citenamefont {Panda}\ \emph {et~al.}(2019)\citenamefont {Panda},
  \citenamefont {Bhagwat}, \citenamefont {Suresh},\ and\ \citenamefont
  {Mitra}}]{regDeconv}%
  \BibitemOpen
  \bibfield  {author} {\bibinfo {author} {\bibfnamefont {S.}~\bibnamefont
  {Panda}}, \bibinfo {author} {\bibfnamefont {S.}~\bibnamefont {Bhagwat}},
  \bibinfo {author} {\bibfnamefont {J.}~\bibnamefont {Suresh}}, \ and\ \bibinfo
  {author} {\bibfnamefont {S.}~\bibnamefont {Mitra}},\ }\href {\doibase
  10.1103/PhysRevD.100.043541} {\bibfield  {journal} {\bibinfo  {journal}
  {Phys. Rev. D}\ }\textbf {\bibinfo {volume} {100}},\ \bibinfo {pages}
  {043541} (\bibinfo {year} {2019})}\BibitemShut {NoStop}%
\end{thebibliography}%
\end{document}